\documentclass[11pt,a4paper]{article}
\usepackage{authblk}
\usepackage{amsmath}
\usepackage{amsfonts}
\usepackage{amssymb}
\usepackage{graphicx}
\usepackage[utf8]{inputenc}
\usepackage{multicol}
\usepackage{graphicx}
\usepackage{subcaption}
\usepackage{float}
\usepackage{wrapfig}
\usepackage[hmarginratio=1:1,
            columnsep=20pt,
            margin=0.8 in,
            top=1in,
            bottom=1in,
            headheight=110pt,
            ]{geometry} 
\usepackage[usenames,dvipsnames,svgnames,table]{xcolor}  
\usepackage[sort&compress,numbers,merge]{natbib}

\usepackage{hyperref} 

\hypersetup{%
    colorlinks=true, linktocpage=true, pdfstartpage=15, pdfstartview=FitV,%
    breaklinks=true, pdfpagemode=UseNone, pageanchor=true, pdfpagemode=UseOutlines,%
    plainpages=false, bookmarksnumbered, bookmarksopen=true, bookmarksopenlevel=1,%
    hypertexnames=true, pdfhighlight=/O,
    urlcolor=Cerulean, linkcolor=Cerulean, citecolor=Cerulean, 
}

\begin{document}

\title{Interplay of Scalar and Fermionic Components in a Multi-component Dark Matter Scenario}
\author[1]{Sreemanti Chakraborti\thanks{sreemanti@iitg.ac.in}}
\author[1]{Poulose Poulose\thanks{poulose@iitg.ac.in}}

\affil[1]{Department of Physics, Indian Institute of Technology Guwahati, Assam 781039, India.}

 \maketitle

\begin{abstract}
We explore the multi-component dark matter (DM) scenario 
considered in a simple extension of the standard model with an inert scalar doublet and a singlet fermionic field providing the two DM candidates. The DM states are made stable under the unbroken $Z_2\times Z_2'$ discrete symmetry. An additional gauge singlet scalar field is introduced to facilitate the interaction of the dark fermion with the visible sector.  Presence of a charged fermionic field having the same $Z_2$ charge as that of the inert scalar field allows exploring the dark matter mass regions otherwise disallowed, like in the standard Inert Doublet Model (IDM) scenarios. 
With these arrangements, it is shown that the light DM scenario and the desert region in the intermediate mass range of DM in the standard IDM case can be made compatible with the relic density bounds and direct detection limits. Further, detailed parameter space study is carried out keeping the coexistence of both the scalar and fermionic components in focus, showing that sizable parameter space regions are available for the entire mass range of $10\ \rm GeV \le M_{DM}\le 2000$ GeV.
\end{abstract}



\section{Motivation}
The existence of dark matter (DM), constituting about 27$\%$ \cite{Aghanim:2018eyx} of the total energy content of the universe, is supported by different independent cosmological observations like galactic rotation \cite{Feng:2010gw,Bergstrom:2012fi,Bertone:2004pz}, the phenomenon of gravitational lensing \cite{Feng:2010gw,Bergstrom:2012fi,Bertone:2004pz},  inhomogeneities in the cosmic microwave background radiation (CMBR) as precisely measured by the WMAP\cite{Komatsu:2014ioa} and the Planck \cite{Aghanim:2018eyx} experiments. On the other hand,
searches for direct observation of the presence of DM has so far produced null results, providing upper bounds on the cross sections of the DM particles scattering off heavy nuclei \cite{Aprile:2017iyp}. While spin dependent and spin-independent cross section measurements are being performed, such direct detection experiments are insensitive to the number of components and the type of dark matter. Weakly Interacting Massive Particle(WIMP) is a popular choice for thermal dark matter candidate, which, being in the mass range of the order of a few GeV to TeV range, can theoretically provide the correct observed relic density and explain the origin of the relic through the thermal freeze-out mechanism. The Standard Model (SM) of particle physics, by design, is silent about dark matter candidates, providing one of the clear reasons to extend the dynamics of the elementary particles beyond SM. Simplest extensions of SM to include DM candidates introduce additional scalar fields, made stable with the help of discrete $Z_2$ symmetry. The DM candidates in these models connect with the visible sector through Higgs portal couplings. Direct detection through nuclear scattering in these models limits these portal couplings of the DM candidates to very small values. Excepting for a very limited parameter space regions, the DM annihilation cross sections with such small couplings are too small to provide the required DM relic density. For a recent review on the singlet scalar dark matter, please see Ref.~\cite{
Athron:2017kgt, Cline:2013gha}. Extending the singlet scalar models to include gauge interactions of the inert scalars is considered in models like the Inert Two Higgs Doublet Model (IDM) \cite{LopezHonorez:2006gr,Hambye:2009pw,Honorez:2010re,Belyaev:2016lok,Goudelis:2013uca, Poulose:2016lvz}, and further extensions in Inert Three Higgs Doublet Models \cite{Chao:2012re,Keus:2013hya}. The gauge interactions of the DM candidates present in such models provide sufficient annihilation cross sections to contain the otherwise overabundant case. In fact, in almost all these cases, the cross section being dependent on the gauge coupling overdo this to bring down the relic density below the required value. Turning to the fermionic DM candidates, the single component framework is studied in, for example, Ref. \cite{Kim:2008pp, Baek:2011aa}. Similar to the case of singlet scalar models, here too, the same Higgs portal interactions decide on the DM-nuclear scattering relevant to the direct detection and the DM annihilation to the visible sector, making the model viable only in a very limited parameter region.
Models going beyond the single component framework are studied within the scalar DM scenarios \cite{Bian:2014cja,DiFranzo:2016uzc, Chiang:2015fta,Bhattacharya:2013hva,Berlin:2016eem, Biswas:2013nn, Bhattacharya:2016ysw, Bhattacharya:2017fid}, however providing only limited distinguishing features compared to the single component framework. 
More recently, serious attempts are made to unify such DM models with features to explain small neutrino mass, another compelling reason to consider beyond-the-Standard Model (BSM) dynamics \cite{Fraser:2015mhb,Ma:2012gb,Bhattacharya:2013nya,Babu:2007sm,Cao:2007fy, Borah:2017dfn, Ma:2006km}.  
 Among these, models with singlet scalar and fermions \cite{Cao:2007fy} face with the similar over abundance problem as that of the corresponding single component frameworks mentioned above.
Adding fermionic fields in the dark sector (odd under the $Z_2$ considered) could provide additional annihilation channels and other possibilities in the DM dynamics, leading to distinctions with the more simplistic scenarios of scalar DM models mentioned above. Studies of simple scenarios with a vector-like dark fermionic field added to the scalar dark matter models show negligible effects in the parameter space regions compatible with the measurements \cite{Borah:2017dqx,Fedderke:2014wda}. Models with  vector and fermion dark matter cases were discussed in Ref.\cite{Ahmed:2017dbb}.

In this article we propose a new scenario with one scalar and one fermionic dark matter particle coexisting to fulfill the relic density conditions, at the same time evading the direct detection possibilities so far.  The key to our new proposal is the identification of the fact that, the annihilation cross section has to be enhanced beyond what is provided by the Higgs portal interactions (which are constrained by the direct detection experiments), at the same time with a handle on the cross section provided by a tuneable coupling, unlike the case of annihilations enabled by the gauge couplings. This, in the proposed scenario, is achieved through the presence of a newly introduced fermionic field, which, along with the standard leptons share a Yukawa coupling with the dark-scalar doublet field. We shall show, that for sufficiently large ranges of the relevant Yukawa coupling and the mass of the new fermion, which are the two new parameters here, it is possible to have the desired relic density for the dark matter candidate. 
These new fermions should carry the same $Z_2$ charge as that of the scalar DM, and therefore are required to be heavier than the DM itself. Thus, DM candidates with larger than about 100 GeV mass, require large values of the Yukawa couplings to compensate for the the scaling down of the cross section with the correspondingly larger $Z_2$-odd fermions, which mediate the annihilations. Thus, it is still desirable to have non-singlet scalar fields to provide the DM candidates. In addition to supporting to obtain the required relic density, the presence of such fermionic partners (considered here as electrically charged) provide new handle to explore this scenario in collider experiments. However,  the gauge mediated annihilations overkill the DM, as in the cases available in the literature. The presence of fermionic partner and the additional channels of annihilation only worsens the situation. The introduction of another DM candidate not only salvages the situation, but in a beautifully interlinked coexistence, provide sufficient relic density in a very large range of masses of both the DM candidates. We demonstrate this in the proposed model through the addition of a gauge singlet fermion field, stable under a different $Z_2'$ symmetry. 
The fermionic dark matter interacts with the SM particles  through a singlet scalar portal,
 enabled by mixing of this neutral scalar with the SM Higgs field,  leading to a natural way to explain direct detection limits through the smallness of the mixing without requiring fine tuning of the parameters of the Lagrangian. The same portal coupling allows interaction with the scalar dark matter candidate field. In fact, such portal interaction between the two dark matter fields enables the conversion of one type of dark matter to the other. In kinematically allowed phase space  regions having a mass hierarchy with the fermionic dark matter heavier than the scalar one, a favourable condition arises with the help of such conversion processes. Specifically, this allows annihilation of the otherwise over-abundant fermionic dark matter to inject the otherwise under-abundant scalar dark matter to provide together the desired relic density. This minimal two-component scalar-fermion dark matter scenario provides interesting additional features as bonus, some of which are detailed in the rest of this article.  

We organise the article with Section~\ref{model} presenting the details of the model, and the features of relic density calculations.We then present our numerical analysis in Section~\ref{analyses}, and summarise the study with our conclusions in Section~\ref{summary}.

\section{Model}
\label{model}
The framework of the proposed model has the same gauge group as that of the SM. 
The particle content of the SM along with the Higgs doublet ($\Phi_1$) is extended with the addition of a scalar doublet $\Phi_{2}$ having hypercharge $+1$, two vector-like fermion singlets, $\chi$ and $\psi$ with hypercharges $-2$ and $0$, respectively, and a singlet scalar field $\phi$, with zero hypercharge. The new doublet field, $\Phi_2$ and charged singlet fermion $\chi$ are considered odd under a discrete $Z_2$ symmetry, while all other fields are considered even under this transformation. Similarly, the neutral singlet fermion $\psi$ is taken to be odd under another $Z_2'$ symmetry, while all other fields are considered even under this.

\begin{figure}
\centering
\includegraphics[scale=0.5]{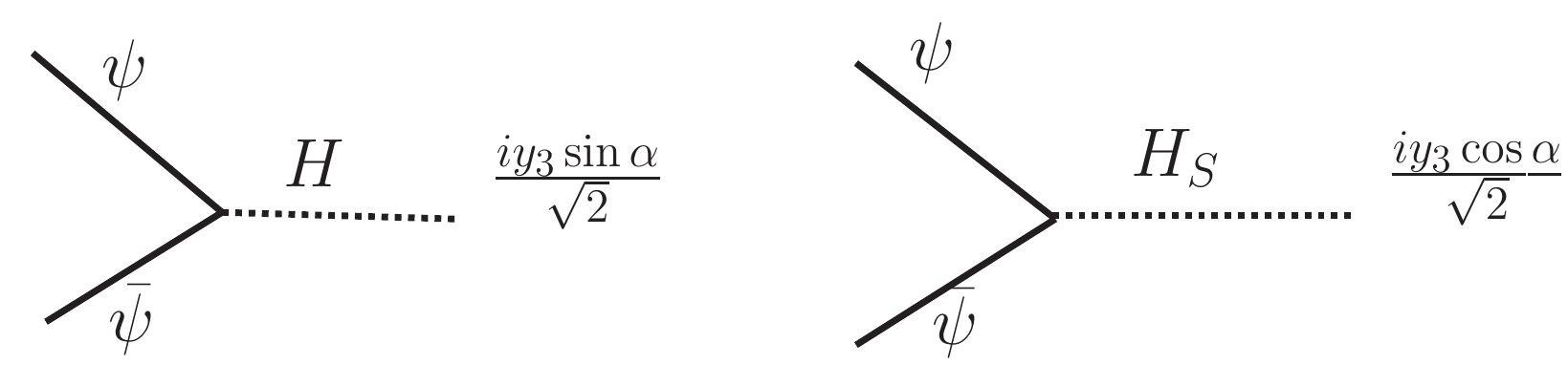}\\
\centering
\includegraphics[scale=0.5]{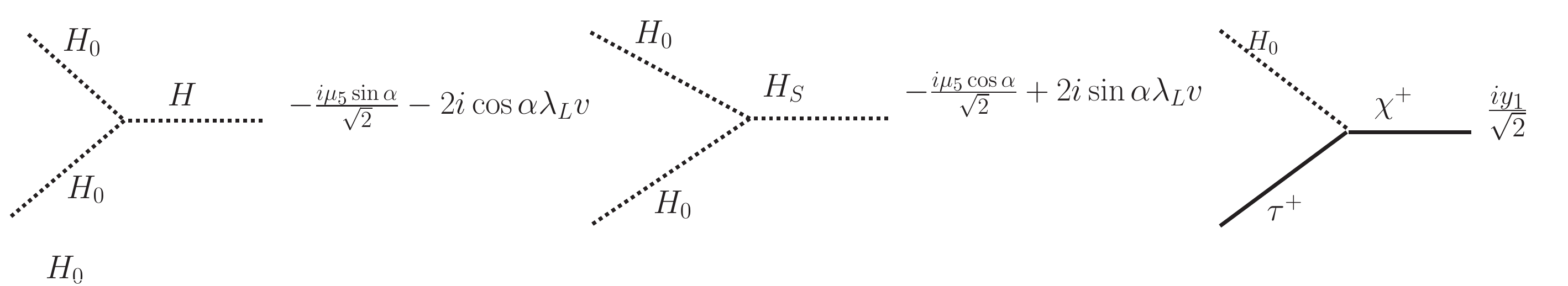}\\
\caption{Feynman rules relevant to the annihilation of the DM candidates.} 
\label{fig_vertices}
\end{figure}

With the above $Z_2\times Z_2'$ discrete symmetry and the SM gauge symmetry, the new physics interaction part of the Lagrangian is given by
\begin{eqnarray}
\mathcal{L} \supset &&\left(D_{\mu}\Phi_{2}\right)^{\dagger}(D^{\mu}\Phi_{2}) +\bar{\chi}i\gamma^{\mu}D'_{\mu}\chi +\bar{\psi}i\gamma^{\mu}\partial_{\mu}\psi- M_{\chi\ }\bar{\chi}\chi - M_{\psi}\ \bar{\psi}\psi \nonumber \\
&&-\big(y_1\ \bar{L}\Phi_{2}\chi_R +h.c.\big)- y_2\ \bar{\chi}\chi\phi - y_3~\bar{\psi}\psi\phi
-V
\end{eqnarray}
with the covariant derivatives $D^\mu=\partial^\mu+ig\tau\cdot W^\mu+ig' \frac{Y}{2}B^\mu$ and $D'^\mu=\partial^\mu+ig'\frac{Y}{2}B^\mu$, where $g$ and $g'$ are the corresponding gauge couplings and $Y$ is the hypercharge. $L$ denotes the SM lepton doublet field.
The scalar potential is given by
\begin{eqnarray}
V&=&\mu_{1}^2\Phi_{1}^{\dagger}\Phi_{1}+\mu_{2}^2\Phi_{2}^{\dagger}\Phi_{2}+\lambda_1(\Phi_{1}^{\dagger}\Phi_{1})^2+\lambda_2(\Phi_{2}^{\dagger}\Phi_{2})^2+\lambda_3( \Phi_{1}^{\dagger}\Phi_{1})( \Phi_{2}^{\dagger}\Phi_{2})\nonumber\\
&&+ \lambda_{4}|\Phi_{1}^{\dagger}\Phi_{2}|^{2}
+\frac{1}{2}\left[\lambda_5(\Phi_1^\dagger\Phi_2)^2+h.c\right] +\mu_{3}^2\ \phi^{\dagger}\phi+ \lambda_6 (\phi^\dagger\phi)^2 
\nonumber \\ 
&&+\frac{1}{2}\left[\mu_4\  \phi (\Phi_1^\dagger\Phi_1)+\mu_5\  \phi (\Phi_2^\dagger \Phi_2) + h.c\right] 
+\frac{1}{2}\left[\mu_6\ \phi^3+\mu_7^3\ \phi+ \mu_8\ (\phi^\dagger\phi)\phi + h.c\right]\nonumber \\
&&+ \lambda_7 (\phi^\dagger\phi)(\Phi_1^\dagger\Phi_1)+\lambda_8 (\phi^\dagger\phi)(\Phi_2^\dagger\Phi_2).
\end{eqnarray} 
With the standard Higgs field developing a vacuum expectation value (vev), $v=246$ GeV, leading to the electroweak symmetry breaking (EWSB), the scalar fields may be expressed in the unitary gauge as
\begin{equation}
\Phi_1=\left(\begin{array}{c}0  \\ \frac{v+h }{\sqrt{2}}\end{array}\right),~~~~\Phi_2=\left(\begin{array}{c}H^+  \\ \frac{H_0+i~A_{0} }{\sqrt{2}}\end{array}\right),~~~~\phi =\frac{1}{\sqrt{2}}(h_s+i~A_s).
\end{equation}
In order to keep the $Z_2$ symmetry intact, we disallow $\Phi_2$ from developing a vev by setting $\mu_2^2 \ge 0$.    Similarly, it is arranged so that $\phi$ does not generate a vev. The physical spectrum now has two charged scalars, $H^\pm$, one neutral scalar $H^0$ and a neutral pseudoscalar $A^0$ coming from $\Phi_2$, with the lightest of $H_0$ and $A_0$ becoming a dark matter candidate. We confine to the case of $M_{H_0}<M_{A_0}$. On the other hand, $h$ and $h_s$ mix to generate the two physical scalar bosons;  the observed 125 GeV Higgs boson and another scalar boson denoted here by $H$ and $H_S$, respectively. This mixing is parametrised with an angle $\alpha$ as
\begin{equation}
\left(\begin{array}{c}
H \\ H_S
\end{array}\right)
= \left(\begin{array}{c c}
\cos\alpha & \sin \alpha \\
-\sin\alpha & \cos\alpha
\end{array}\right)\left(\begin{array}{c}
h \\ h_s
\end{array}\right).
\end{equation}
The physical scalar masses are related to the quartic coupling $\lambda_1$ and the vev through the relation
\begin{equation}
{M_{H_S}^2-M_{A_S}^2+M_H^2}={2\lambda_1v^2}.
\label{eq_lambda1}
\end{equation}
Considering $\lambda_1$ to be positive for the stability of the potential leads to the tree-level mass relation
\begin{equation}
{M_{H_S}^2+M_H^2 > M_{A_S}^2}.
\end{equation}
Mixing with the SM Higgs field allows the scalar component of the singlet field $H_S$ to decay to the SM particles, thus allowing its mass to be practically unrestricted. On the other hand the pseudoscalar component, $A_s$ directly couples only to the the new fermions and the inert Higgs field, and thus to allow tree-level decay its mass is required to be larger than twice the mass of the lightest dark matter candidate.  
The mass spectrum of the inert doublet field are not affected by other interactions, with the masses related to the parameters of the potential as in the pure IDM case given by \cite{LopezHonorez:2006gr}
\begin{eqnarray}
M_{H_\pm}^2 &=& \mu_2^2+\frac{\lambda_3}{2}~v^2\nonumber \\
M_{H_0}^2 &=& M_{H_\pm}^2+\frac{1}{2}~(\lambda_4+\lambda_5)~v^2\nonumber \\
M_{A_0}^2 &=&M_{H_\pm}^2+\frac{1}{2}~(\lambda_4-\lambda_5)~v^2.
\label{eq_mass}
\end{eqnarray}
In addition, as explained in the introduction, the new scenario necessitates two additional charged leptons,  and a neutral fermion $\psi$ in the physical spectrum. The mass hierarchy of $M_{\chi^\pm}> M_{H_0}$ is maintained to allow the decay of  $Z_2$ odd fermion, $\chi^\pm$, whereas $\psi$, the fermionic component of the dark matter is made stable with the $Z_2'$ symmetry.
Apart from the mass relations in Eq.~\ref{eq_mass} above, the condition 
\begin{equation}
\mu_7^3 =-\frac{\mu_4}{2}~v^2
\end{equation} 
is set to remove the linear term after the EWSB. This along with setting $\mu_3^2,~\mu_6,~\mu_8,~\lambda_6$ and $\lambda_7$  to be positive definite makes sure that $\phi$ does not develop a non-zero vev. In our analysis we have traded the parameter $\mu_3^2$ for the physical mass, $M_{A_S}$, which are related through
\begin{equation}
\mu_3^2= M_{A_S}^2-\frac{\lambda_7 v^2}{2}.
\end{equation}
In our choice of parameters, we have made sure that $\mu_3^2\ge 0 $, as required. We list the vertices and the corresponding Feynman rules relevant to the new degrees of freedom in Fig.~\ref{fig_vertices}.
Further, we define  
\begin{equation}
\lambda_L=\frac{1}{2}(\lambda_3+\lambda_4+\lambda_5)
\end{equation} 
as the combination of the couplings that is relevant to Higgs portal interaction of the scalar dark matter candidate involved in the direct detection experiments as well as the annihilation processes. Coming to the experimental constraints,  LEP limits of 
\begin{equation}
(M_{H_0}+M_{A_0},~2M_{H^+}) > M_Z; ~~~~{\rm and}~~~~~M_{H_0}+M_{H^+} > M_W
\end{equation}
are obtained from non observation  of the decay of  $Z$ and $W$ to the inert Higgs bosons. The precision electroweak measurements are sensitive to the mass splitting between the charged Higgs boson and the neutral ones, with the IDM contribution to the $T$-parameter given by
\begin{equation}
T_{IDM}=\frac{1.08}{v^2}~(M_{H^\pm}-M_{H_0})(M_{H^\pm}-M_{A_0}).
\end{equation}
The current experimental bound on the value of $T=0.08\pm0.12$ \cite{Tanabashi:2018oca} can be accommodated with at least one of the light neutral Higgs bosons having mass close to that of the charged Higgs boson.

\begin{figure}
\centering
\includegraphics[scale=0.5]{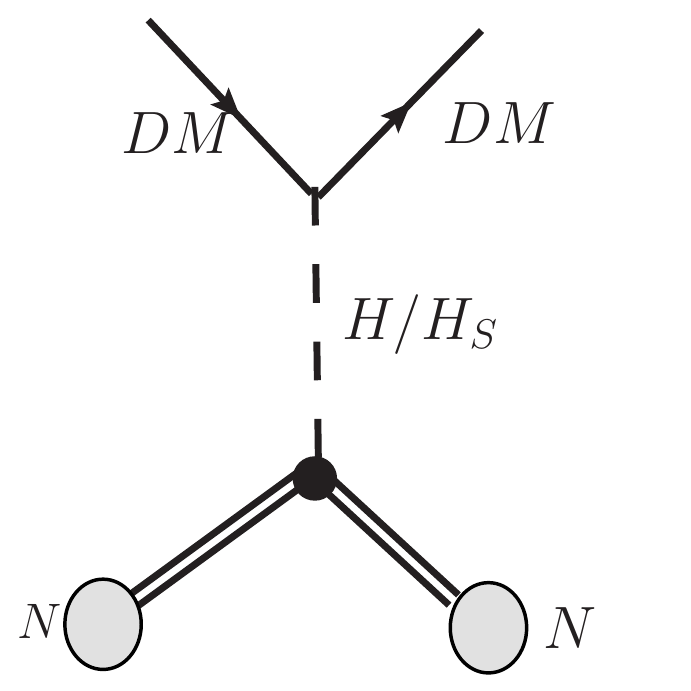}
\caption{Schematic diagram showing the scattering of DM with nucleus considered in the direct detection experiments.} 
\label{fig_dd}
\end{figure}

The direct detection depends on the elastic scattering of the dark matter candidate with the neutron and proton in the nucleus, which is mediated by the scalar bosons in our case, as shown in Fig.~\ref{fig_dd}. This does not, therefore, distinguish whether the dark matter is a fermion or a scalar particle. While the  scalar dark matter candidate has a direct coupling with the SM doublet field, the fermionic dark matter interacts with the visible sector only by virtue of the mixing between the new scalar field introduced and the standard Higgs boson. The scattering cross section of the scalar DM is dictated by the coupling $\lambda_L$ in the IDM sector, and the newly introduced trilinear coupling $\mu_5$ between $\phi$ and the inert doublet field $\Phi_2$. The process is mediated by $H$ and $H_S$ with the former case coupling to $H_0$ with appropriate combination of $\lambda_Lv\cos\alpha$ and $\mu_5 \sin\alpha$, and the latter case with combination of $\lambda_Lv\sin\alpha$ and $\mu_5\cos\alpha$, as could be read from Fig.~\ref{fig_vertices}. The relevant spin independent cross sections are given by

\begin{eqnarray}
\sigma_{H_0N \rightarrow H_0N}&=&\frac{f^2_N~M_N^4}{16\pi(M_{H_0}+M_N)^2}\left(\frac{\lambda_{H}}{M_H^2}~\cos\alpha-\frac{\lambda_{H_S}}{M_{H_S}^2}~\sin\alpha\right)^2,  \nonumber \\
 \sigma_{\psi N \rightarrow \psi N}&=& \frac{f^2_N~M_N^4}{16\pi(M_{\psi}+M_N)^2}\left(\frac{ g_{H}}{M_H^2}~\cos\alpha-\frac{g_{H_S}}{M_{H_S}^2}~\sin\alpha\right)^2,
\label{DD_xsection}
\end{eqnarray}
where 
\begin{eqnarray}
\lambda_{H}&=& -\frac{1}{\sqrt{2}}\left(\mu_5 \sin\alpha-2\sqrt{2}~\lambda_L v~\cos\alpha\right),\nonumber \\
\lambda_{H_S}&=&-\frac{1}{\sqrt{2}}\left(\mu_5\cos\alpha+2\sqrt{2}~\lambda_Lv~\sin\alpha\right),\nonumber \\
g_ {H}&=&\frac{y_3}{\sqrt{2}}~\sin\alpha,~~~~~~~~~~~ \noindent\\
g_ {H_S}&=&\frac{y_3}{\sqrt{2}}~\cos\alpha,
\end{eqnarray}
and, $ f_N$ is the nuclear form factor. Notice that in the absence of mixing ($\sin\alpha=0$), $\sigma_{H_0N \rightarrow H_0N}$ is reduced to the usual case in the pure IDM~\cite{Barbieri:2006dq}.  Further, $\sigma_{\psi N \rightarrow \psi N}$ is proportional to $(\sin\alpha\cos\alpha)^2$, reminding us that $\psi$ interacts with the visible sector owing entirely to the mixing of $\phi$ with the doublet Higgs field.

\begin{figure}{H}
\centering
\includegraphics[width=0.48\linewidth]{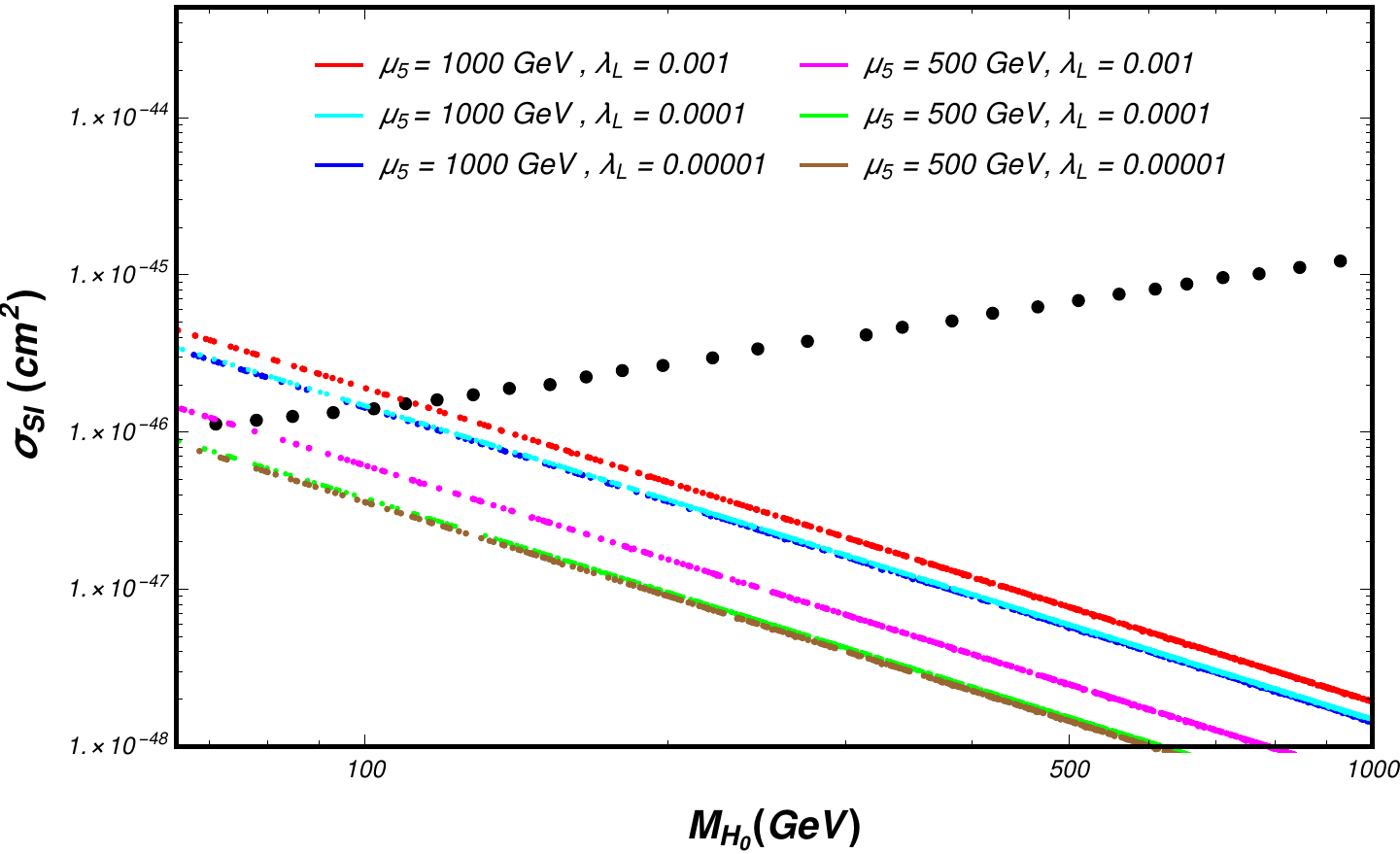}
\hfill
\includegraphics[width=0.48\linewidth]{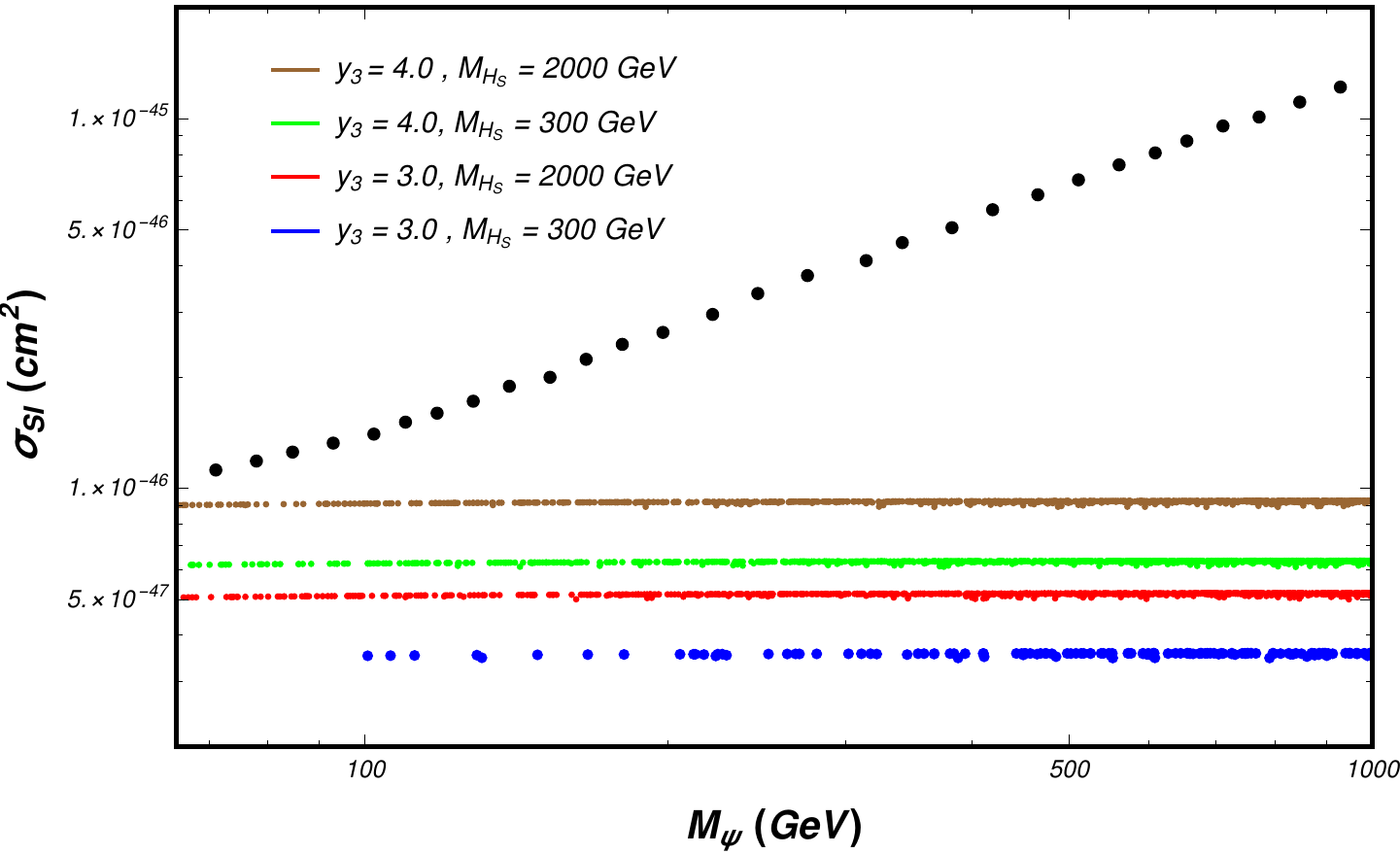}
\caption{$\sigma_{SI}$  vs $M_{DM}$ for different couplings and mediator mass relevant to direction detection processes. The black dotted lines represent the current XENON IT bound\cite{Aprile:2018dbl}.}
\label{direct-limits}
\end{figure}

To find the viable parameter values, we compare the direct detection  cross section obtained using micrOMEGAs~\cite{Belanger:2014vza}  with the  XENON1T \cite{Aprile:2018dbl} bounds.
In Fig.~\ref{direct-limits} (left) we plot the cross section for $H_0$-nucleon scattering against $M_{H_0}$ for two different values of $\mu_5=$ 500 GeV and 1000 GeV, with three different values of $\lambda_L=10^{-3},~10^{-4},~10^{-5}$ in each case.  As seen, the sensitivity of $\lambda_L$ is insignificant for sizable $\mu_5$. While we have presented our results in Fig.~\ref{direct-limits} for slightly larger values of $\mu_5$, a similar pattern is seen in the case of smaller values ($\mu_5\sim 100$ GeV) as well.  Dependence on $M_{H_S}$ is not presented, however, we have checked that it is not very significant. Notice that $\mu_5 \le 500$ GeV is compatible with $M_{H_0} \ge 65$ GeV, and $\mu_5 \le 1000$ GeV is compatible with $M_{H_0}\ge 100$ GeV.  While considering the limits, we have not included the contribution of the other dark matter candidate, $\psi$, which means the scaling factor in 
\( \frac{\Omega_i}{\Omega_{\rm tot}}  \sigma_i \) \cite{Herrero-Garcia:2017vrl} for the i$^{\rm th}$ DM component in multi-component scenario is taken as 1. Moving on to the case of $\psi$ (again, in the absence of $H_0$), cross section of $\psi$-nucleon scattering is plotted against $M_\psi$ for different values of the $\psi\psi\phi$ coupling $y_3$ in Fig.~\ref{direct-limits} (right). Here again the mediators are $H$ and $H_S$. However, the couplings are rather straight forward, unlike the previous case of $H_0$-nucleon scattering, with the cross section proportional to $(y_3\sin2\alpha)^2$ irrespective the case with $H$ or $H_S$ as the propagator, as clear from Eq.~\ref{DD_xsection} and discussion there.  We have considered a fixed value of $\alpha=0.0045$ and varied $y_3$ to see the effect of the coupling. Apart from the coupling, the process depends strongly on $M_{H_S}$, the mediator mass. As can be seen from Fig.~\ref{direct-limits} (right), $y_3=4$ is allowed for the above value of mixing, and a rather heavy $H_S$ with $M_{H_S}=2$ TeV. Lighter $H_S$ leads to more relaxed limit on the coupling, contrary to the naive expectation. This could be attributed to the destructive interference between the contributions from $H_S$ mediation and $H$ mediation. With these observations, we proceed to see the effect on the dark matter relic density.

In order to understand the compatibility of the model with the observations of relic density given by $\Omega\rm h^2$ = 0.1198 $\pm$ 0.0012\cite{Aghanim:2018eyx}, where $h$ denotes the Hubble parameter normalised to 100 km $s^{-1}\rm Mpc^{-1}$, we perform a relic density computation using the micrOMEGAs package, scanning over the theoretically available parameter space regions. We consider two distinct scenarios with (i) $M_{H_0} \le \frac{M_H}{2}$ and (ii) ~$M_{H_0} > \frac{M_H}{2}$.  In the former case, the invisible decay of the Higgs boson to DM pairs will put restrictions on the couplings. Here, 
$\lambda_L$ is the relevant coupling for  $H\rightarrow H_0 H_0$ process, although, through the $H$-$H_S$ mixing, this channel is also influenced by the the trilinear coupling, $\mu_5$ and the mixing angle, $\alpha$. Considering the fermionic component of the DM, notice that the only interaction of $\psi$ to other particles is facilitated by the singlet scalar field $\phi$. Through mixing with the SM Higgs field, this leads to $\psi\psi H$ coupling of $y_3\sin\alpha$. The present LHC bound on $H\rightarrow {\it invisible}$ decay width is restricted to about 20\% \cite{Albert:2017onk}, leading to a constraint on $y_3 \le 0.02 $ for the maximum allowed $\alpha\sim 0.33$.
In our analysis, in the region $M_{H_0}\le \frac{M_H}{2}$ and $M_{\psi}\le \frac{M_H}{2}$, we have discussed two scenarios. One with $\lambda_L=\mu_5=y_3=0$ as the most conservative approach, so that the invisible decay of $H$ to the dark matter particles is disallowed, and the second case in which we relax this with non-zero values of these couplings, which are compatible with the present experimental limit on the invisible decay of Higgs boson.
However, in the first case setting $y_3$ to be zero whenever $M_\psi \le \frac{M_H}{2}$ results in over abundance of $\psi$ and consequently ruling out these mass ranges for $\psi$. Therefore, in those discussions we shall consider $M_\psi > \frac{M_H}{2}$.
Finally, the Yukawa coupling $y_1$ between the SM leptons and dark fermion $\chi$ with the inert scalar field, when allowed for the first two generations may induce larger than desired $(g-2)$ for electron and muon. In addition, the presence of non-diagonal couplings can induce undesired lepton-flavour-violating processes like $\mu\rightarrow e\gamma$. To avoid such effects,  we consider the couplings to be diagonal and $\chi$ that couples to the first two generations to be very heavy, along with suitably chosen small values of the corresponding couplings. Therefore, in our further discussion we consider only the third generation coupling to be present, leading to $\chi\tau H_0$ interaction. As is clear from the discussion below, this coupling and the corresponding interaction makes significant impact only in the {{light $H_0$ case, ie, 45 GeV $\le M_{H_0} \le$ 80 GeV .}} Before embarking on our numerical analysis, we shall look at the details of the above scenarios.

\subsection{Scenario 1: $M_{H_0}\le \frac{M_H}{2}$}\label{sec_ov_lowmass}

\begin{figure}[H]
\hspace*{5mm}
\begin{subfigure}{.4\textwidth}
\centering
\includegraphics[trim=0 -40 0 0,clip,width=0.50\linewidth]{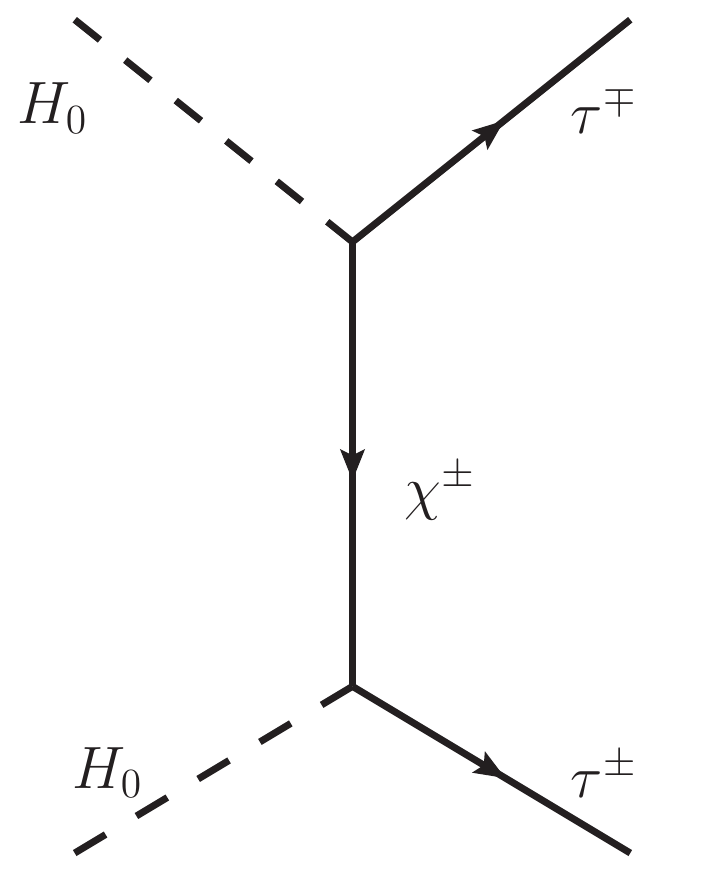}
\caption{}
\label{fd_H0H02tautau}
\end{subfigure}
\hspace*{5mm}
\begin{subfigure}{.4\textwidth}
\centering
\includegraphics[scale=0.4]{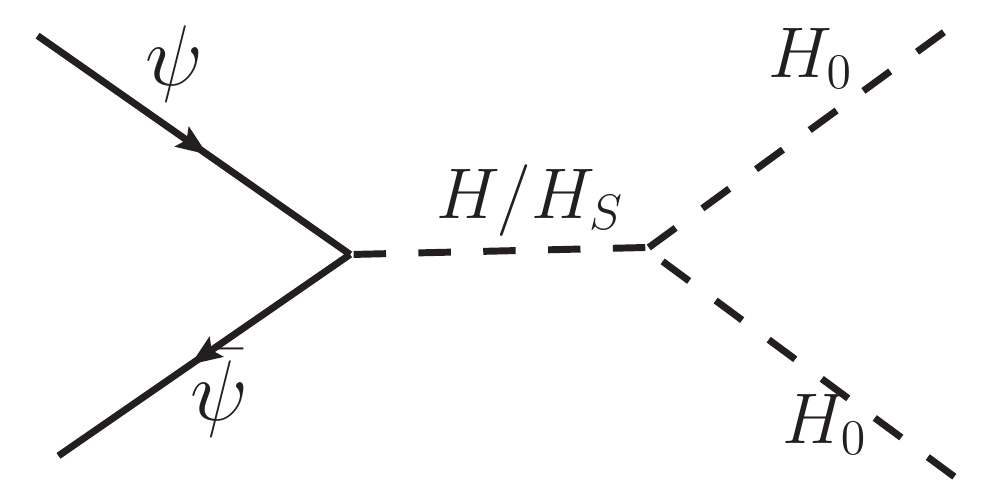}
\caption{}
\label{fd_psipsi2H0H0}
\vfill
\centering
\includegraphics[scale=0.45]{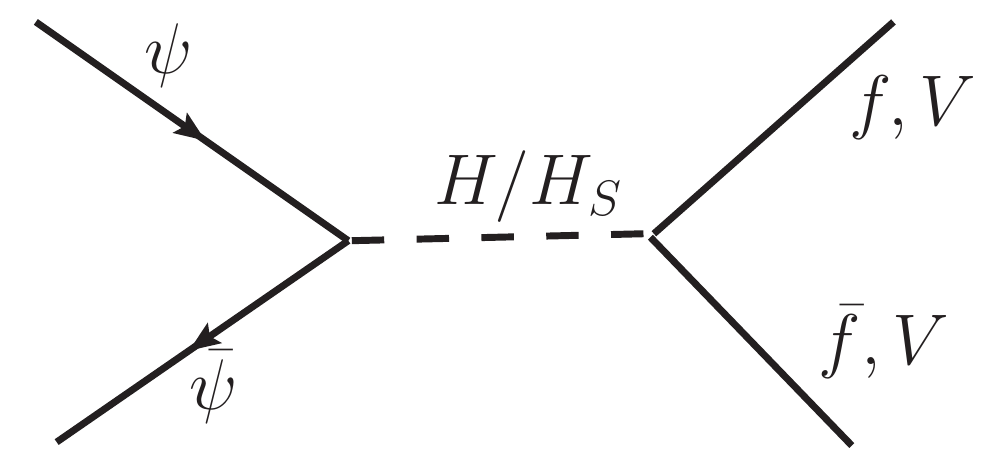}
\caption{}
\label{fd_psipsi2SMSM}
\end{subfigure}
\caption{Feynman Diagram showing (a) the $t$-channel annihilation channel of $H_0$, 
(b) the conversion of $\psi$ to the scalar DM candidate and (c) the annihilation of $\psi$ into SM final states, where $V=W,~Z$}
\label{figure_FD1}
\end{figure}

Keeping all other scalar masses larger than its mass, $H_0$ annihilates into the SM leptons and light quarks, mediated through $H$ and $H_S$ in the $s$-channel, and through the newly introduced heavy fermion, $\chi$ in the $t$-channel. Notice that, in the absence of $\chi$ the annihilation is enabled through the $s$-channel process, $H_0H_0\rightarrow (H,~H_S)\rightarrow f\bar f$, where $f$ is the SM fermion. The cross section of this process is proportional to 
\begin{equation}
\left(\frac{\lambda_{H}}{M_H^2}~\cos\alpha-\frac{\lambda_{H_S}}{M_{H_S}^2}~\sin\alpha\right)^2,
\nonumber
\end{equation}
the same coupling factor appearing in the direct detection cross section in Eq.~\ref{DD_xsection}, and, therefore, stringently  constrained. In that case, the cross section is not sufficient to bring down the relic density to the observed value. Thus, the presence of an additional scalar singlet does not help improve the situation.The presence of $\chi$, however, changes the scenario by adding to the cross section a $t$-channel process $H_0H_0\rightarrow \tau\tau$, mediated by $\chi$. The corresponding Feynman diagram is given in Fig.~\ref{fd_H0H02tautau}, and the cross section is given in Eq. \ref{h0h0tautau}.
This additional cross section can be tuned with the help of the unrestricted $H_0 \chi \tau$  Yukawa coupling $y_1$ to get the required relic density. On the other hand, for large values of $y_1$ the annihilation of $H_0$ can make it underabundant. 
With $M_\psi > M_{H_0}$, the contribution to relic density from the $H_0$, denoted by $\Omega_1\rm h^2$, is independent of $M_\psi$ itself, except for a small dependence on the conversion process $\psi\psi \rightarrow H_0H_0$. 
This additional creation of $H_0$ is controlled by the trilinear and quartic couplings $\mu_5$ and $\lambda_L$, the mass of $H_S$, and the Yukawa coupling, $y_3$, as is clear from the Feynman diagram in Fig.~\ref{fd_psipsi2H0H0}. 
The cross section for this process is given in the Appendix, Eq.~\ref{psipsih0h0}.
Setting $\lambda_L=\mu_5=0$ to avoid invisible decay of $H$, as stated above, makes this channel irrelevant. 
In addition to $\Omega_1\rm h^2$,  the total relic density($\Omega_{\rm tot}  \rm h^2$) has the fermionic component, $\Omega_2  \rm h^2$, so that $\Omega_{\rm tot}  \rm h^2=\Omega_1  \rm h^2+\Omega_2 \rm h^2$. Note that $\Omega_2  \rm h^2$ is controlled by the $\psi$ annihilation into the SM states. 
Before the gauge boson annihilation channels open up for $\psi$ at $M_\psi\sim 80$ GeV, the fermionic component $\Omega_2  \rm h^2$ is larger than the allowed relic density, unless the other annihilation channel $\psi\psi \rightarrow ff$, where $f$ denotes the SM fermions, is sizable. This latter process, mediated by the singlet component of $H$ and $H_S$ depends on the combination of $y_3 \sin2\alpha$ through the $(\psi\psi H,~ffH)$ pair of interactions (see Fig.~\ref{fig_vertices}) in the $H$ mediated case, and through $(\psi\psi H_S,~ffH_S)$ pair in the $H_S$ mediated case. The Feynman diagram for this process is given in Fig.~\ref{fd_psipsi2SMSM}, and cross section in Eq.~\ref{psipsiff}.
 On the other hand, for $M_\psi \ge 80$ GeV, the $\psi\psi \rightarrow VV$, where $V=W,~Z$, channel allows considerable reduction in $\Omega_2  \rm h^2$, opening large parameter space region compatible with the current measurements. The cross section for these processes is given in Eq.~\ref{psipsiWW} and \ref{psipsiZZ}. While these processes are also suppressed by the Higgs mixing, these are much more significant compared to the $f\bar f$ annihilation channel with
 \begin{equation}
 \frac{\sigma_{\psi \bar{\psi}\rightarrow V V}}{\sigma_{\psi \bar{\psi}\rightarrow f\bar f}}=\frac{1}{x~N_C}\frac{M_V^4}{4M_\psi^2 m_f^2}\frac{\beta_V}{\beta_f^3}\left(3-\frac{4M_\psi^2}{M_V^2}+\frac{16M_\psi^4}{M_V^4}\right),
 \end{equation}
 where $x=8$ for $V=W$ and $x=16$ for $V=Z$, and the color factor, $N_C=1$ for leptons and $N_C=3$ for quarks. As these gauge annihilation channels are $s$-channel processes mediated by $H$ and $H_S$,  the allowed parameter region is expected to be around the resonant condition $M_{H_S} \sim 2 M_\psi$. Away from this region, for $M_\psi \ge  M_{H_S}$, the possibility of $\psi$ annihilation is controlled by the $\psi\psi\rightarrow H_SH_S$ process.  While the $H/H_S$ mediated $s$-channel depends on the  trilinear couplings $\mu_6$ and $\mu_8$,  the more important $t$-channel (see Fig.~\ref{fig_psipsi2hshs} for the Feynman diagram) depends on  the Yukawa coupling $y_3$. The cross section for this process is given in Eq. \ref{psipsihshs}. There is also a less relevant $\psi\psi \rightarrow HH$ possibility, which, however, is proportional to $(y_3\sin\alpha)^2$.

\paragraph{\it Interplay of the DM components:}    Note that, the individual components should be in an underabundant state so that the total relic density is within the desired bound. 
A situation that warrant particular attention is the case of $\frac{M_H}{2} < M_\psi <80 $ GeV. Here, the annihilation of $\psi$ is decided mainly by two processes leading to $ff$  and $H_0H_0$ final states.  
 In addition, when $M_\chi < M_\psi$ it is possible to have pair of $\psi$ annihilating into a pair of $\chi$. Feynman diagram for this process is given in Fig.~\ref{fig_psipsi2chichi}, and  the corresponding cross section is given in Eq.~\ref{psipsichichi}.
 Since the conversion channel depends on the couplings which are constrained by the direct detection experiments and the invisible Higgs decay bounds, the presence of $\chi$ is necessary to have the combined relic density in the allowed limit.
For non-zero values of $\lambda_L,~\mu_5$ and $y_3$, the only substantial change is in $\Omega_2\rm h^2$, which could now be reduced to within the observed bound even for $M_{\psi} < M_V$ due to the additional $H$ and $H_S$ mediated channels.

\subsection{Scenario 2:  $M_{H_0} > \frac{M_H}{2}$}\label{sec_ov_largemass}
In this region $\lambda_L$ and $\mu_5$ are relatively unconstrained, opening up possibilities beyond what is discussed in 
{\it Scenario 1} with $M_{H_0} \le \frac{M_H}{2}$. This along with the fact that $H_0$ can be lighter than $\psi$ makes the Boltzmann's equations for each of these species interdependent. Thus, $\Omega_1\rm h^2$ would now depend not only on the  mass and couplings of $\psi$, but also on its number density. 
The process $\psi\psi \rightarrow H_0H_0$ is now relevant to both $\Omega_1\rm h^2$ and $\Omega_2\rm h^2$, dictated by $\mu_5,~\lambda_L,~ y_3$ and the mixing angle $\alpha$.  
It may be noted, that with only inert scalar present, a substantial region of parameter space with $80 \lesssim M_{H_0} \lesssim 500$ GeV 
is underabundant due to the large annihilation into the gauge bosons. Now with the presence of $\psi$ and the possibilities mentioned above, it opens up a large window of DM mass region accessible. 

\section{Numerical Results}\label{analyses}

We now come to our numerical results in this section.
The quartic couplings, $\lambda_2$ and $~\lambda_6$ involve  four-point self  interactions, and therefore do not influence the relic density computations. Similarly, $\lambda_7,~\mu_4,~\mu_6$ and $\mu_8$ would influence the annihilation through $s$-channel into singlet scalar or the SM Higgs boson mediated by these same scalar fields. $\lambda_8$, which can involve in the $H_0H_0$ annihilation to the singlet scalars through a four-point interaction on the other hand will not have much influence 
on the analysis. 
We have, therefore,  fixed $\lambda_2=\lambda_6=\lambda_7=\lambda_8=0.1$, and $\mu_4=\mu_6=\mu_8=100$ GeV in our study. This leaves the Yukawa couplings, 
 $y_1,~y_2,~y_3$, the quartic coupling combination $\lambda_L$, and the trilinear coupling between the scalar singlet and the inert doublet, $\mu_5$, apart from the relevant masses, which we consider as independent parameters in our numerical analysis.

\subsection{Scenario 1: $M_{H_0} \leq \frac{M_H}{2}$}
As discussed in Section~\ref{sec_ov_lowmass}, in Case 1 we set $\lambda_L=0$ and $\mu_5=0$ and $M_{\psi} > \frac{M_{H}}{2}$ in this scenario
 and in Case 2 we consider non zero values of these parameters compatible with both invisible Higgs decay width and direct detection bounds. Considering the LEP constraint, we keep $M_{H_0} > 45$ GeV. Other parameters are considered as given in Table~\ref{s1scan1}. This choice corresponds to the quartic couplings of the Lagrangian in the ranges, $\lambda_1 = (0.26,0.56),~\lambda_4=(0.262, 0.335 ),~\lambda_5=(-0.377, -0.297),~\lambda_3=(0.036, 0.042)$.

\begin{table}[h]
\begin{tabular}{c |c  }
\hline
\hline
$M_{H_0}$ & $45\leq M_{H_0} \leq 63$ \\
\hline
$M_{A_0}$ & $M_{H_0}+60$ \\
\hline
$M_{H^{\pm}}$ & $M_{H_0}+65$  \\
\hline
$M_{A_S}$ & $121 \leq M_{A_S} \leq 500$  \\
\hline
$M_{H_S}$ & $M_{A_S},M_{A_S}+50$ \\
\hline 
$M_{\chi}$ & (75, 130, 200) \\
\hline
$y_1,~y_2,~y_3$ & $0 \leq y_i \leq 3$ \\
\hline
$\mu_4,\mu_6,\mu_8$ & 100  \\
\hline
$\lambda_2,~\lambda_6,~\lambda_7,~\lambda_8$ & 0.1  \\
\hline
$\alpha$ & 0.0045\\
\hline  
\end{tabular}
\hspace{0.5 cm}
\begin{tabular}{c |c |c }
\hline
\hline
Case 1 & & Case 2 \\
\hline
$65\leq M_{\psi} \leq 1000$ & $M_{\psi}$ & $10\leq M_{\psi} \leq 1000$ \\
\hline
0 & $\lambda_L$ & 0.0001  \\
\hline
0 & $\mu_5$ &100 \\
\hline
\end{tabular}
\caption{Parameters considered for the DM relic density study in scenario 1. All mass parameters are in GeV.}
\label{s1scan1}
\end{table}
In the case of IDM Higgs bosons, LEP rules out the region where $M_{H_0}<$ 80 GeV, $M_{A_0}<$ 100 GeV and $M_{A_0}-M_{H_0}>$8 GeV, since it would lead to visible di-lepton or di-jet signals \cite{Lundstrom:2008ai}. At the same time, mass splitting below 8 GeV does not support relic bound \cite{Belyaev:2016lok}. We have checked that the situation does not change in the present model. Further, LEP-II constrains $M_{H^{\pm} }>  70$ GeV from non-observation of $e^+ e^- \rightarrow H^+ H^-$ production \cite{Pierce:2007ut}. The Electroweak Precision Measurements (EWPM) require product of the mass splittings, $(M_{H^+}-M_{A_0})(M_{H^+}-M_{H_0})$ to be small \cite{LopezHonorez:2010tb}. These considerations have led to deciding $M_{A_0}$ to be larger than 100 GeV for the range of $M_{H_0}$ considered here, keeping $M_{H^+}$ close to $M_{A_0}$. At the same time, $M_{A_0} > 180$ GeV in this set up would correspond to $|\lambda_{4,5}| >1$. We, therefore, set a mass splitting of $M_{A_0}-M_{H_0}=60$ GeV as our conservative choice. The choice of $M_{A_s} > 2M_{H_0}$ is made to allow tree-level decay of $A_s$.  As indicated by Eq.~\ref{eq_lambda1} we shall keep the mass splitting  between that of $H_S$ and $A_s$ small enough to keep $\lambda_1$ small. At the same time, to keep $\lambda_1$ positive all through the parameter region, we make the conservative choice of $M_{H_S}>M_{A_s}$.

\begin{figure}[h]
\centering
\includegraphics[width=0.7\linewidth]{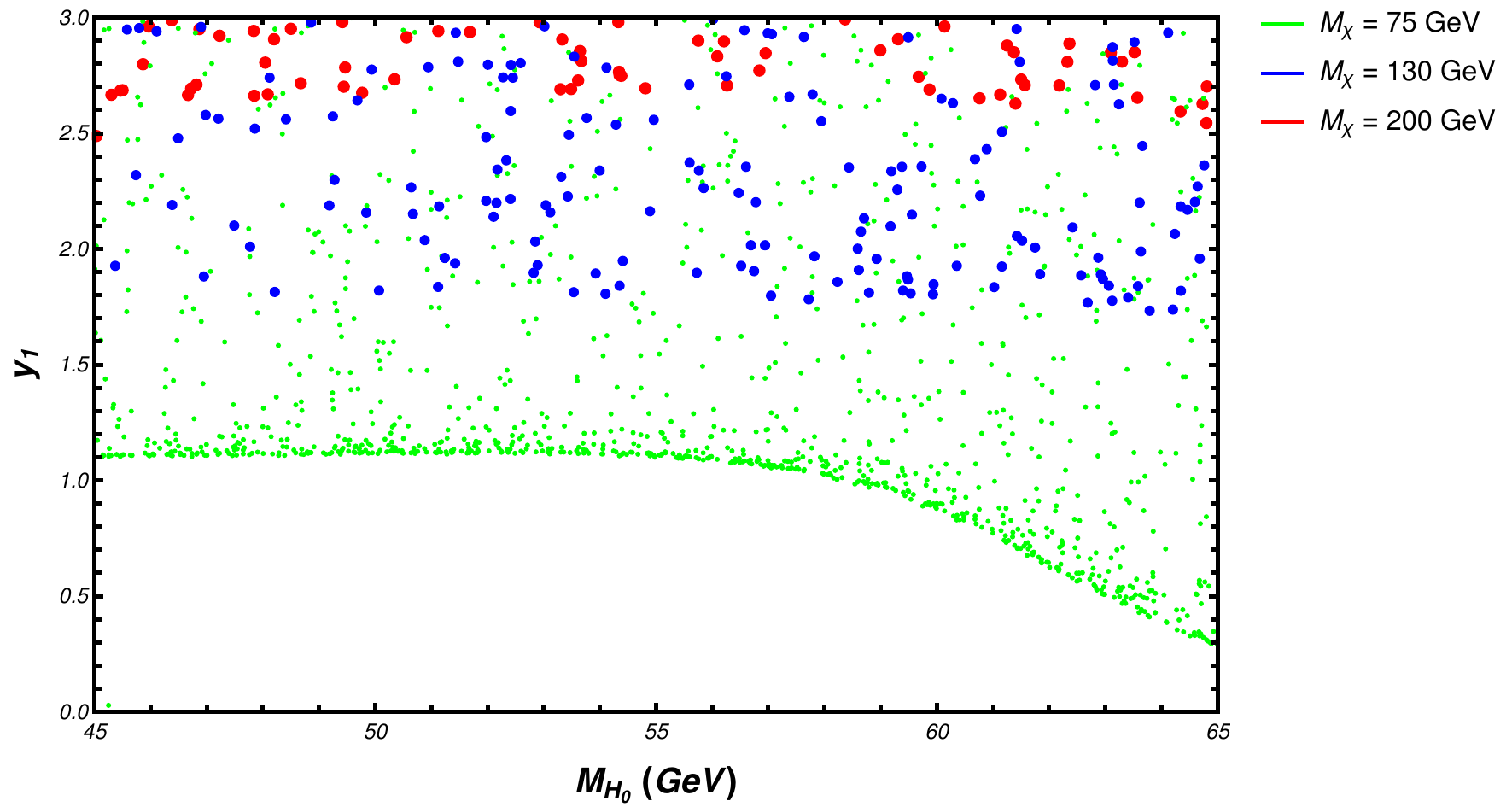} 
\caption{Region of  $y_1-M_{H_0}$ plane compatible with total relic density within the allowed range for selected values of $M_\chi$. Other variables are varied as per Table~\ref{s1scan1}.}
\label{figure_FD1_y1MH0}
\end{figure}

\begin{figure}[H]
\centering
\begin{subfigure}{.55\textwidth}
\includegraphics[scale=0.4]{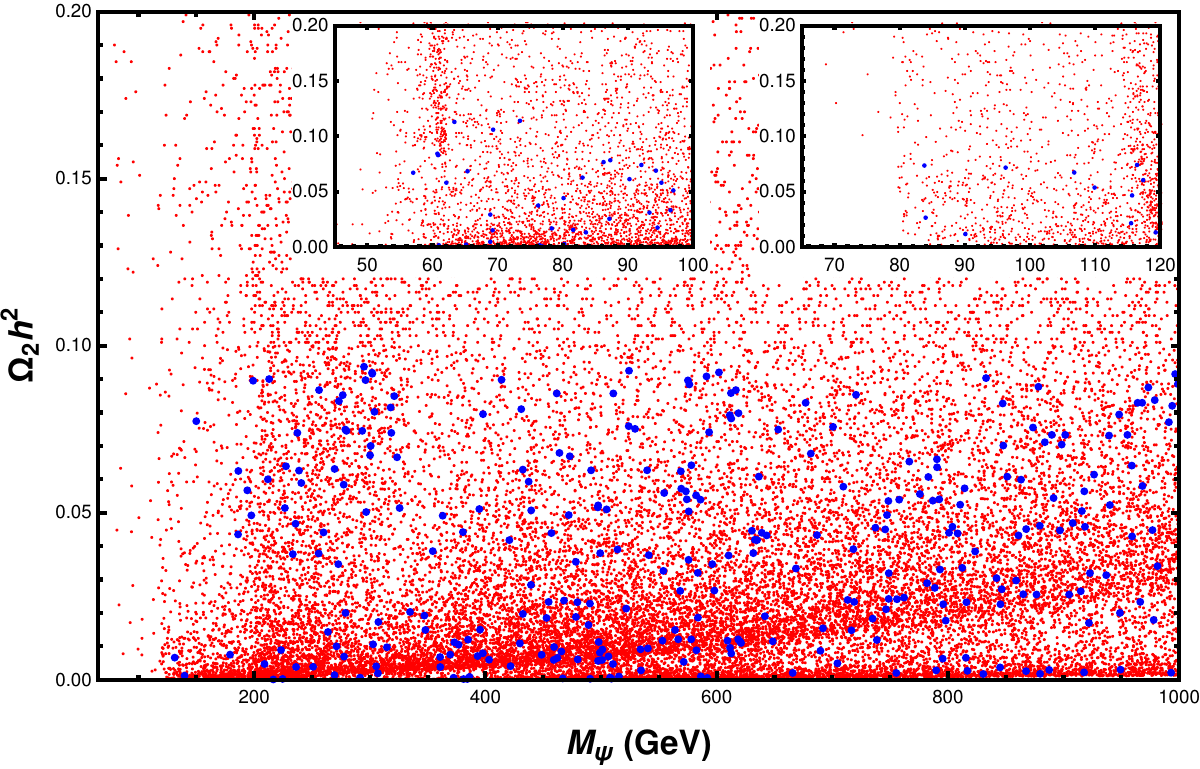}
\end{subfigure}
\caption{$\Omega_2 \rm h^2$ vs $M_\psi$ for the parameter region considered in Table~\ref{s1scan1}. Blue points are a subset satisfying the total relic density $\Omega_{\rm tot}\rm h^2$ within the allowed range.  Small mass region is enlarged for clarity : {\it Top-left inset:} for parameter region Case 2, {\it Top-right inset:} for parameter region Case 1 in Table~\ref{s1scan1} }
\label{figure_Om2Mpsi}
\end{figure}

\begin{figure}[H]
\centering
\begin{subfigure}{.4\textwidth}
\centering
\includegraphics[scale=0.4]{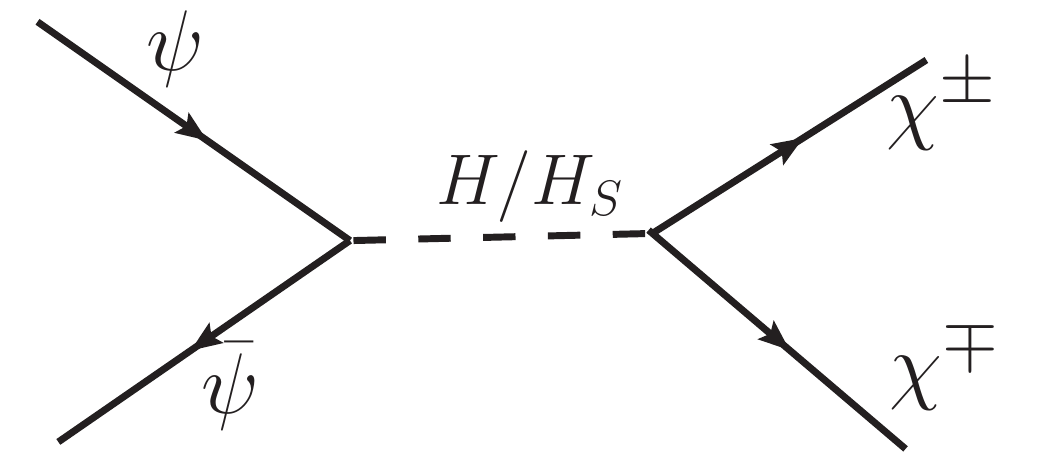}
\caption{}
\label{fig_psipsi2chichi}
\vfill
\centering
\includegraphics[scale=0.45]{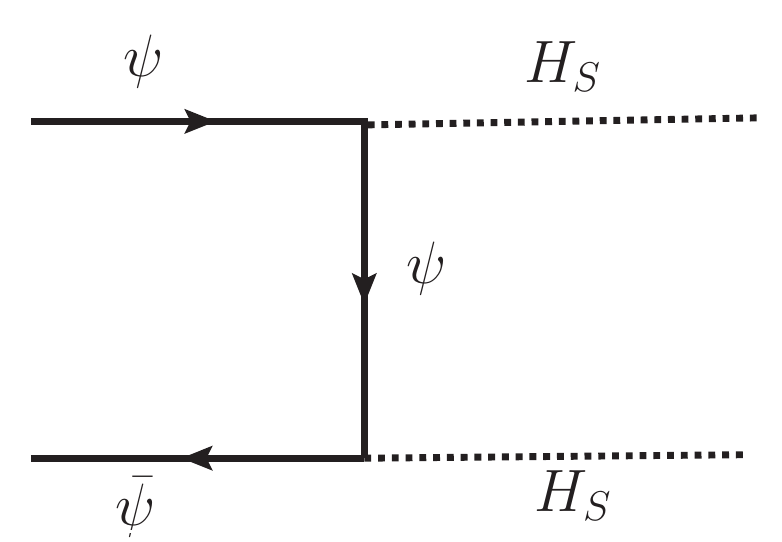}
\caption{}
\label{fig_psipsi2hshs}
\end{subfigure}
 \hskip 5mm
 \begin{subfigure}{.55\textwidth}
\includegraphics[scale=0.4]{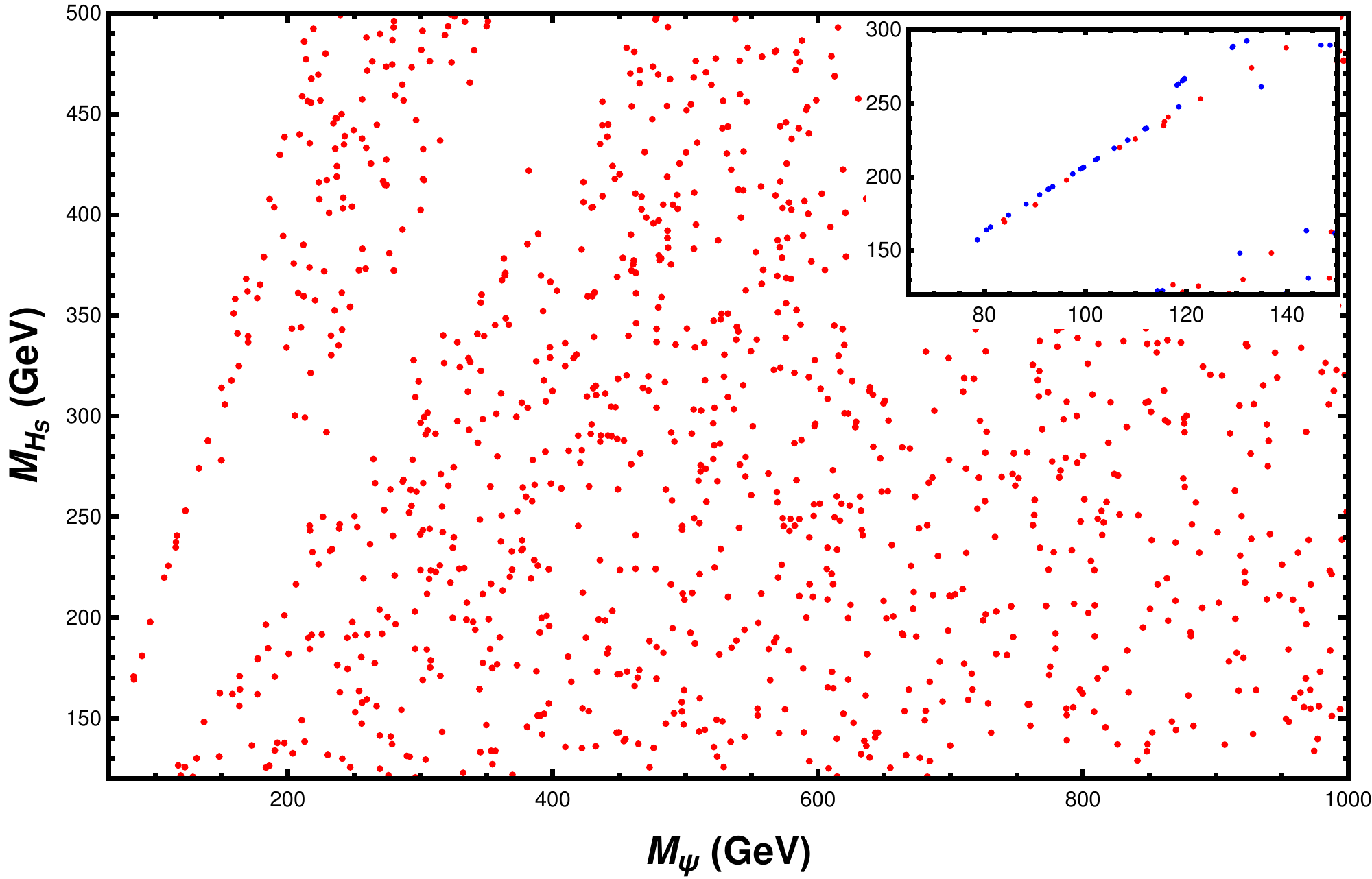}
\end{subfigure}
\caption{({\it Right}): Regions of $M_{H_S}-M_\psi$ with $\Omega_{\rm tot} \rm h^2$ within the allowed relic density range. {\it Inset:} Low mass region is enlarged for clarity. The blue points correspond to $\Omega_2\rm h^2$ alone within the experimental bound. ({\it Left}): Feynman diagrams showing the relevant annihilation channels. }
\label{figure_MH0_mpsi}
\end{figure}

With the above choice of parameters, {micrOMEGAs} is used to perform a random scan to compute the relic density to find compatible regions. With $\lambda_L=0=\mu_5$ disabling all the $H$ and $H_S$ mediated processes, the cross section is dominated by the $\chi$ mediated processes discussed in the previous section. The compatible regions in $y_1 - M_{H_0}$ plane for three different illustrative choices of $M_\chi$ is presented in Fig.~\ref{figure_FD1_y1MH0}. As expected, heavier mediator require larger coupling for the same level of cross section to satisfy the bounds. The other candidate for dark matter, $\psi$ couples directly only to the singlet scalar field, $\phi$, as elaborated in the previous section. Thus, its annihilation process is mediated by $H_S$, whose major component is $\phi$, and the observed 125 GeV resonance of $H$ having a small admixture of $\phi$ enabled by the nonzero mixing angle, $\alpha$. The number density of otherwise over abundant $\psi$ is reduced with the opening of the gauge annihilation channel with $M_\psi \ge M_W$ as clear from the top-right inset of  Fig.~\ref{figure_Om2Mpsi} in the $\Omega_2\rm h^2$ vs $M_\psi$ plot. 
For Case 2 in Table~\ref{s1scan1} of parameter choice where $\lambda_L$ and $\mu_5$ are non-zero, the only visible change in the results is shown in the top left-inset. Here $\Omega_2\rm h^2$ is reduced to the observed bound from overabundance even for $M_{\psi} < M_W$, thanks to the additional annihilation of $\psi$ into SM leptons and $H_0$ now made possible with the non-zero couplings. Since all these channels are $H$ and $H_S$ mediated, we also see the s-channel resonance effect at $M_{\psi}=\frac{M_H}{2}$ in this plot.
$H_S$ dependence in the relic density calculations come through its mediation of $\psi$ annihilation processes as well as annihilation of $\psi$ pairs into $H_S$ pairs. Thus the resonant enhancement of the cross section indicate that the compatible region has $M_{H_S}\sim 2M_\psi$, as clear in Fig.~\ref{figure_MH0_mpsi}. When the full range of $M_{\psi}$ is considered,the annihilation channels of $\psi \psi \rightarrow H_S H_S,~HH,~\chi^{\pm}\chi^{\mp} $ open up.
The mass relation between $M_{\psi}$ and $M_{H_S}$ is no longer linear due to these new channels, leading to  the scattered points in the high mass range starting from 125 GeV. Notice that $\psi \psi \rightarrow H_S H_S$ $t$-channel process and $\psi \psi \rightarrow \chi \chi$ process have negligible dependence on the mixing angle $\alpha$, whereas for all other processes $\sin\alpha$ appears in combination with $y_3$. Thus, in most situations a change in $\alpha$ is compensated by a corresponding change in $y_3$.

\begin{figure}[H]
  \centering
  \includegraphics[width=0.8\linewidth]{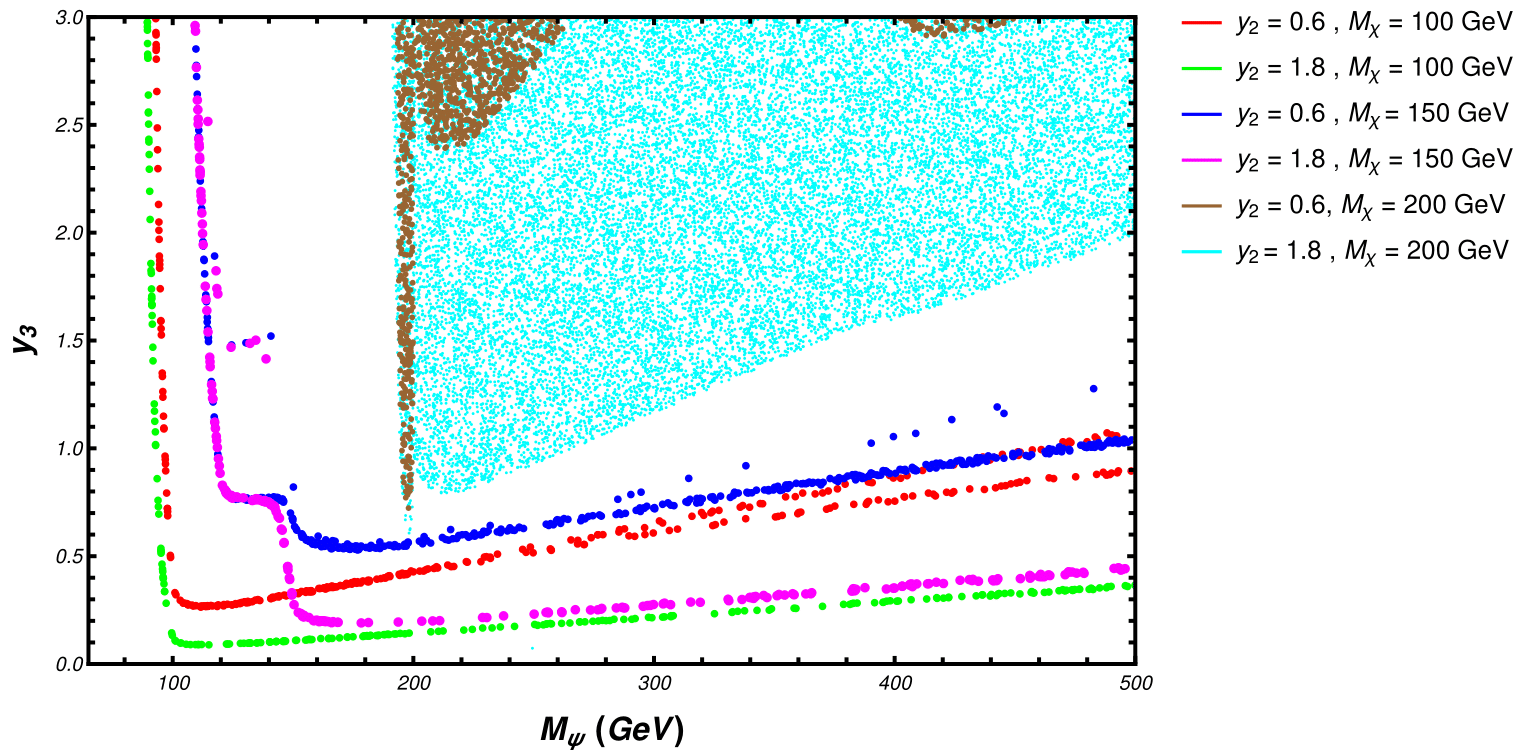}
  \caption{ Regions of $y_3-M_\psi$ plane compatible with $\Omega_{\rm tot} \rm h^2$ satisfying the experimental bounds, for specific values of $y_2$ and $M_\chi$.}
    \label{y3Mpsi}
\end{figure}

For $M_\psi \ge M_\chi$ the $s$-channel $\psi$ annihilating into $\chi$ process is mediated by the singlet scalar, and is proportional to the product of the Yukawa couplings, $y_2\cdot y_3$. In Fig.~\ref{y3Mpsi} regions on $y_3-M_\psi$ plane compatible with the total relic density bound is shown for  $M_\chi=100\ \rm GeV,~150\ \rm GeV$ and $200$ GeV, in each case for two different choices of $y_2 $ values. $M_\psi=M_\chi$ threshold is clearly seen in all the cases considered. Further, for small values of $M_\chi$, the contribution from $\Omega_1\rm h^2$ to the relic density is negligible, as the $\chi$ mediated $t$-channel cross section being large washes it out, minimising the spread in the allowed region. However, for $M_\chi=200$ GeV and above, $\Omega_1  \rm h^2$ has non-negligible contribution, as clearly indicated in the larger area of allowed regions. The dependence on $y_2$ compared to $y_3$ is somewhat trivial as mentioned above. Points with different $y_2$ values shown unambiguously brings out the role of the $\psi\psi \rightarrow \chi\chi$ process. The split lines in the case of smaller $y_2$ values is due to the effects of other processes like $\psi$ pair annihilating into the singlet scalars.

\subsection{Scenario 2: $\frac{M_H}{2} \leq M_{H_0}$} \label{sec_scen2n}

\begin{table}[H]
\centering
\begin{center}
\begin{tabular}{c |c  }
\hline
\hline
$M_{\psi}$ & $65\leq M_{\psi} \leq 1000$  \\
\hline
$M_{H_0}$ & $65\leq M_{H_0} \leq 1000$ \\
\hline
$M_{A_0}$ & $M_{H_0}+(20,~60,~1)$ \\
\hline
$M_{H^{\pm}}$ & $M_{H_0}+(20,~65,~1)$  \\
\hline
$M_{A_S}$ & $131 \leq M_{A_S} \leq 1000$  \\
\hline
$M_{H_S}$ & $M_{A_S}+50$ \\
\hline 
\end{tabular}
\hspace{1cm}
\begin{tabular}{c |c  }
\hline
\hline
$M_{\chi}$ & 1200 \\
\hline
$y_1,~y_2,~y_3$ & $0 \leq y_i \leq 3$ \\
\hline
$\lambda_L$& 0.0001  \\
\hline
$\mu_5$ & (0, 100, 200, 500)  \\
\hline
$\mu_4,\mu_6,\mu_8$ & 100  \\
\hline
$\lambda_2,~\lambda_6,~\lambda_7,~\lambda_8$ & 0.1  \\
\hline
$\alpha$ & 0.0045  \\
\hline
\end{tabular}
\caption{Parameters considered for the DM relic density study in scenario 2. All mass parameters are in GeV.}
\label{s2scan1}
\end{center}
\end{table}

With non-zero values of $\lambda_L$ and $\mu_5$, corresponding to the couplings of $\psi\psi \phi$ and $H_0H_0\phi$, respectively, the process $\psi\psi\rightarrow H_0 H_0$ mediated by $\phi$ (Fig.~\ref{fd_psipsi2H0H0}) makes the interaction between the two dark matter components more relevant. The set of coupled Boltzmann's equations make $\Omega_1\rm h^2$ dependent directly on $\Omega_2\rm h^2$ and vice versa. The value of $HH_0H_0$ coupling is kept at $\lambda_L=10^{-4}$, so as to respect the direct detection limit. Four different illustrative values including zero is considered for $\mu_5$.

\begin{figure}[H] 
\centering
  \includegraphics[width=0.3\linewidth]{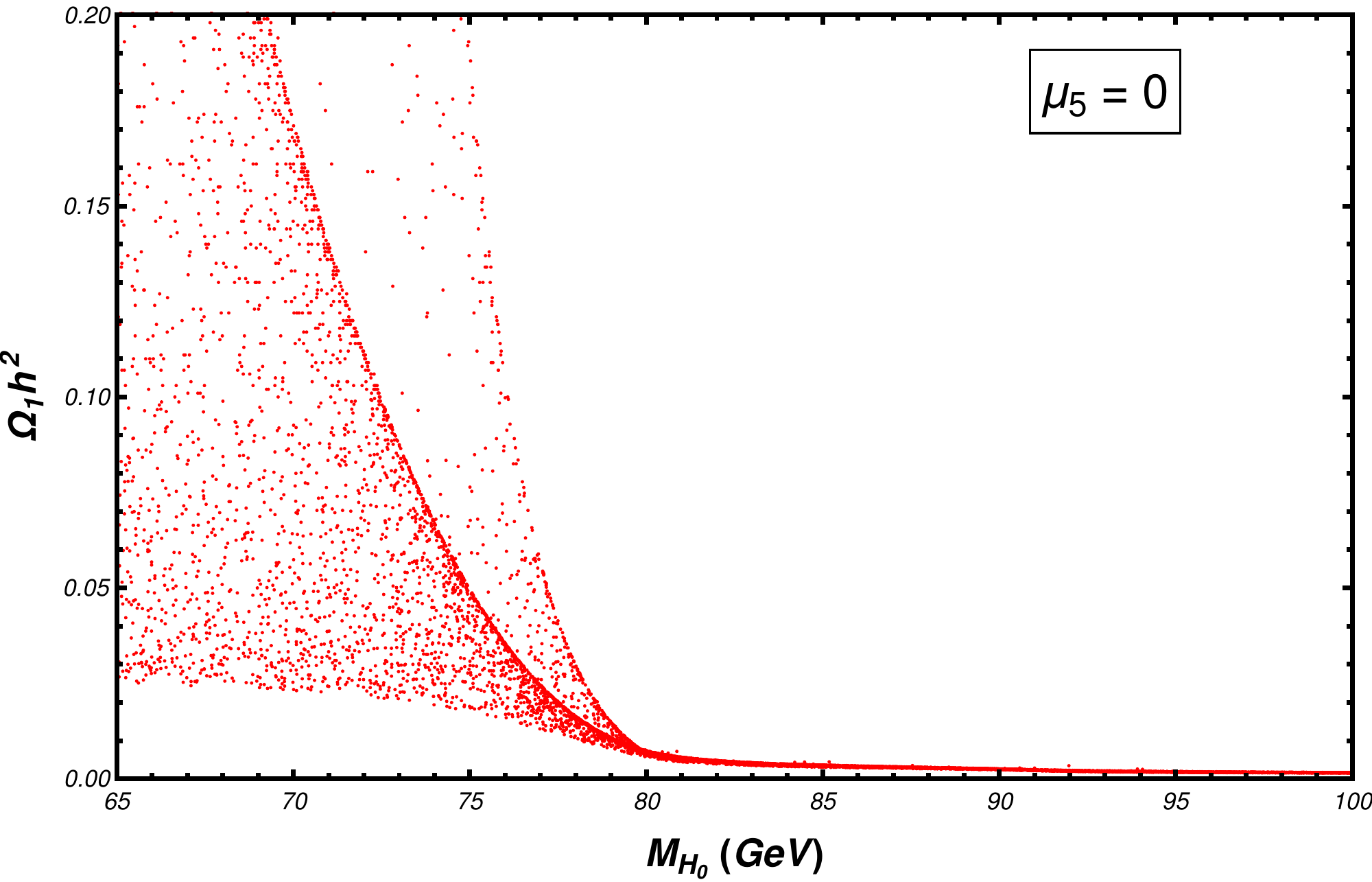}
  \hfill
  \includegraphics[width=0.3\linewidth]{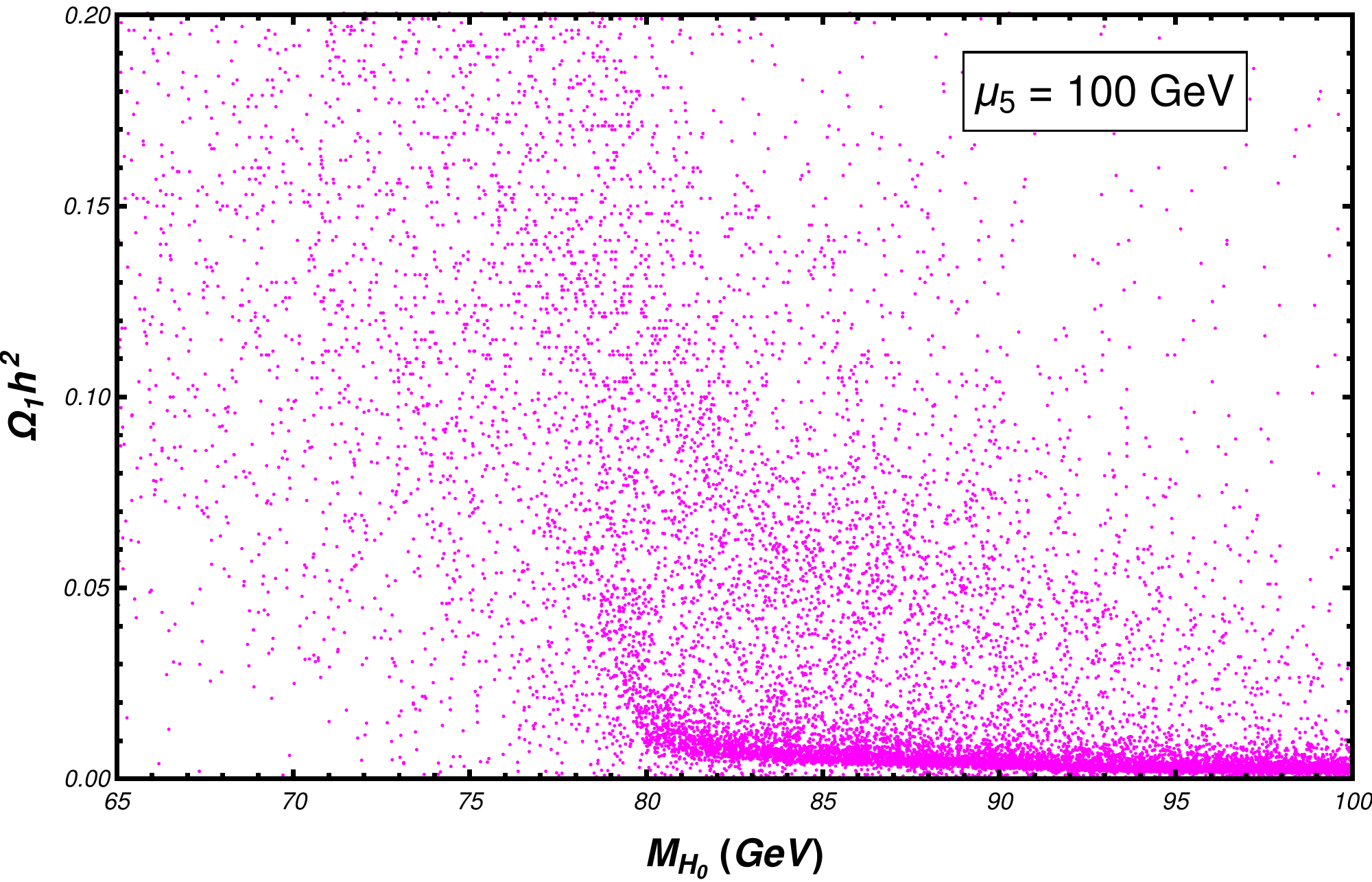}
  \hfill
  \includegraphics[width=0.3\linewidth]{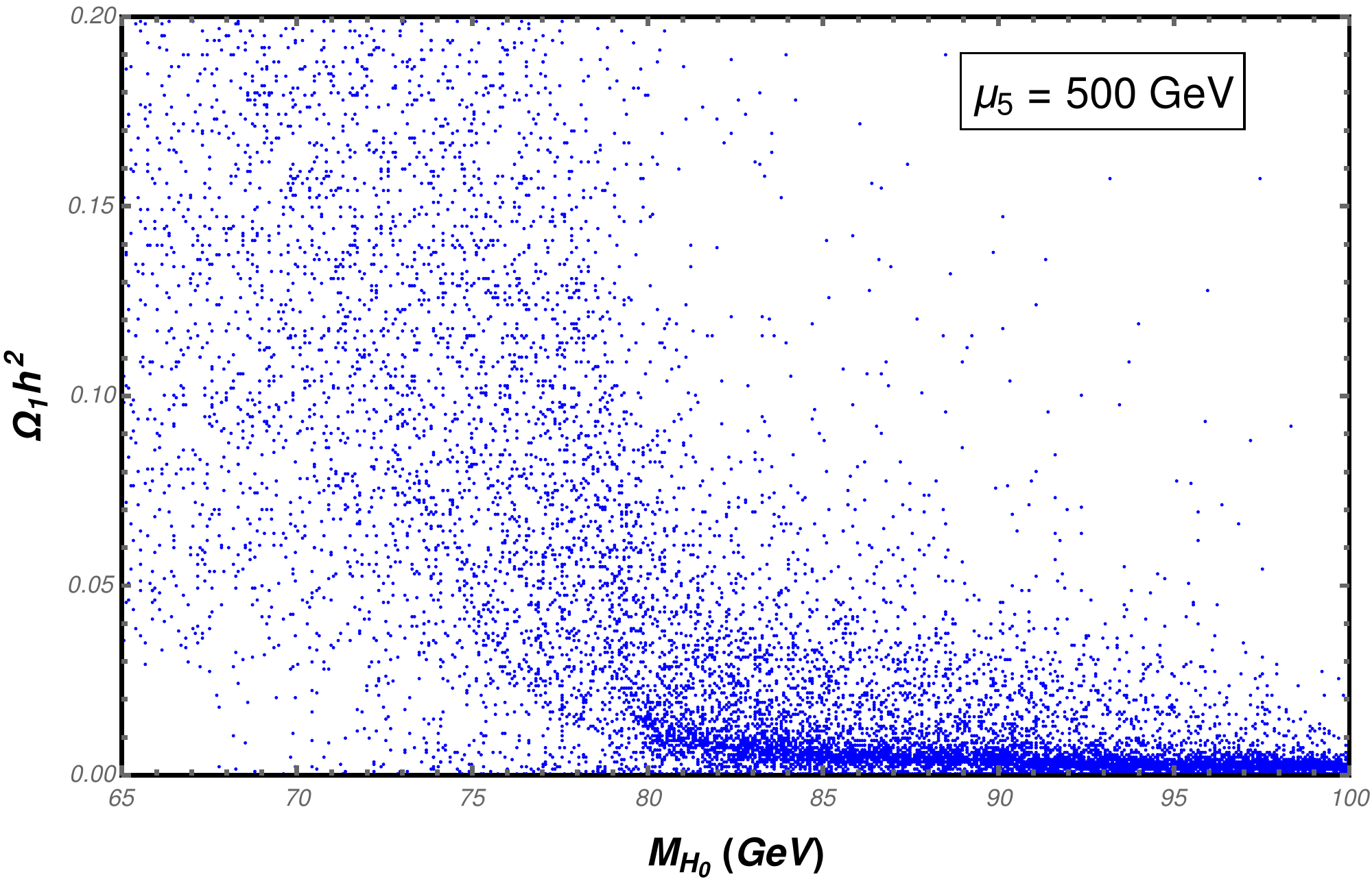}\\
  \bigskip
  \includegraphics[width=0.3\linewidth]{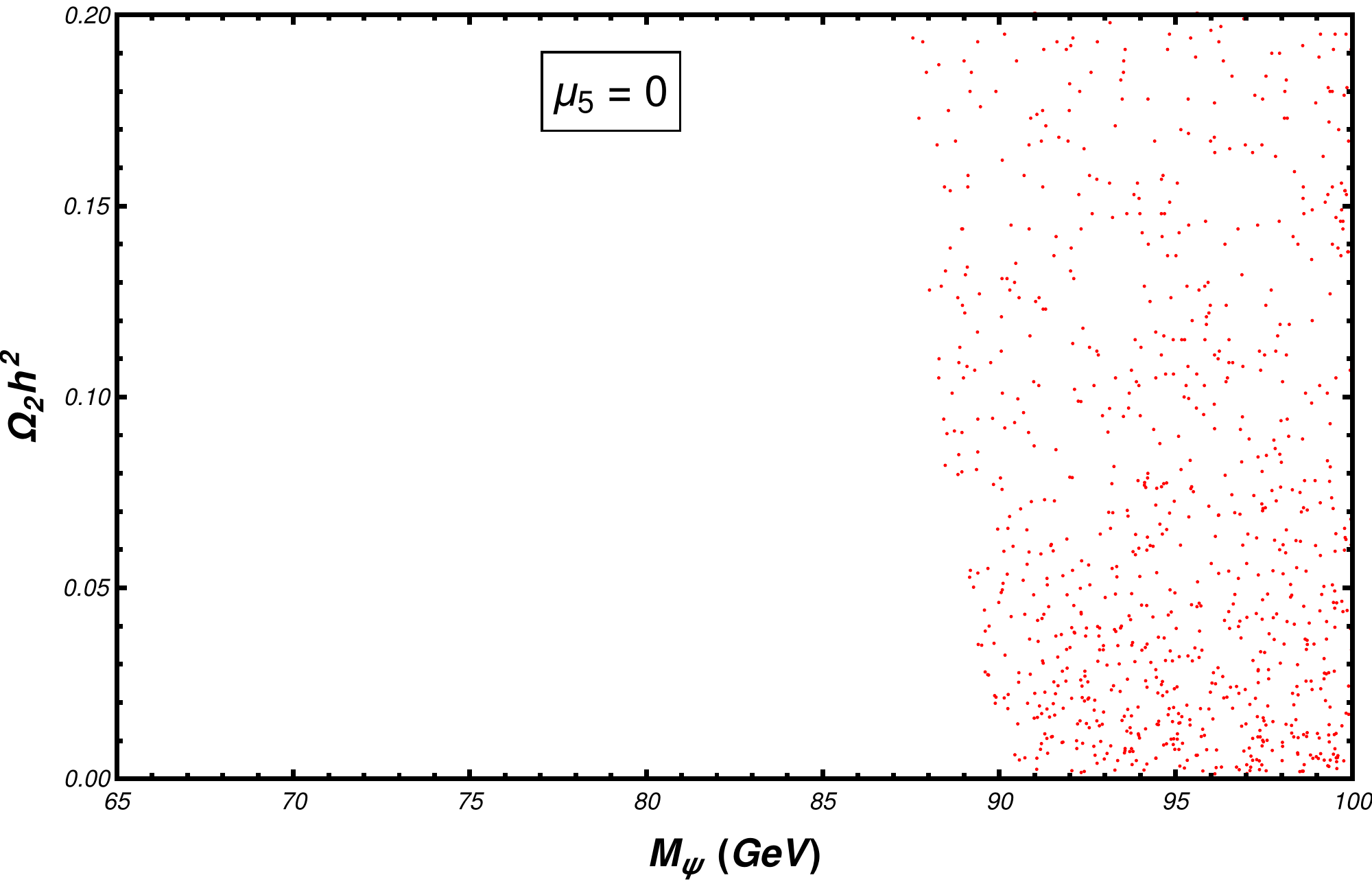}
  \hfill
  \includegraphics[width=0.3\linewidth]{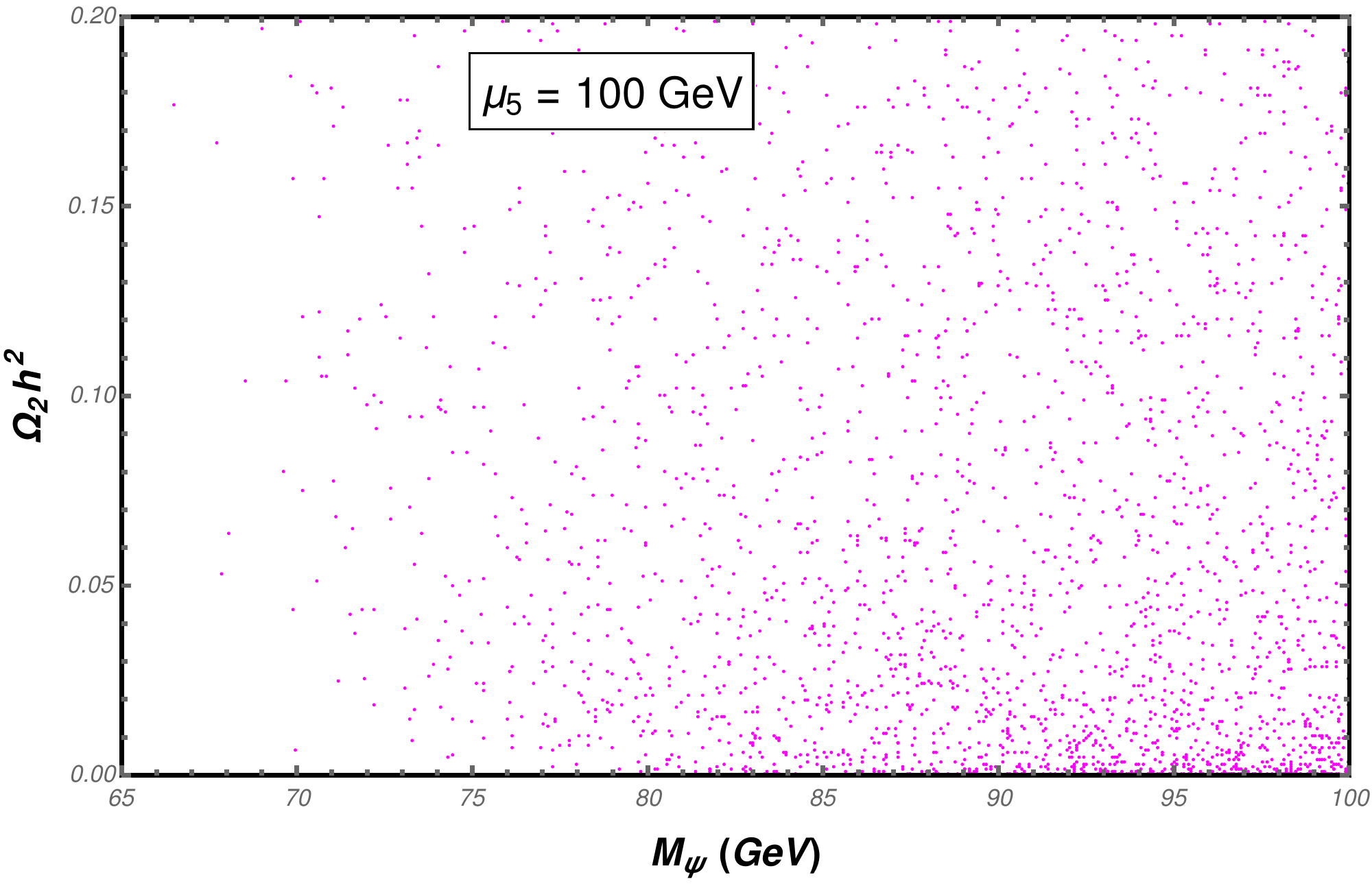}
  \hfill
  \includegraphics[width=0.3\linewidth]{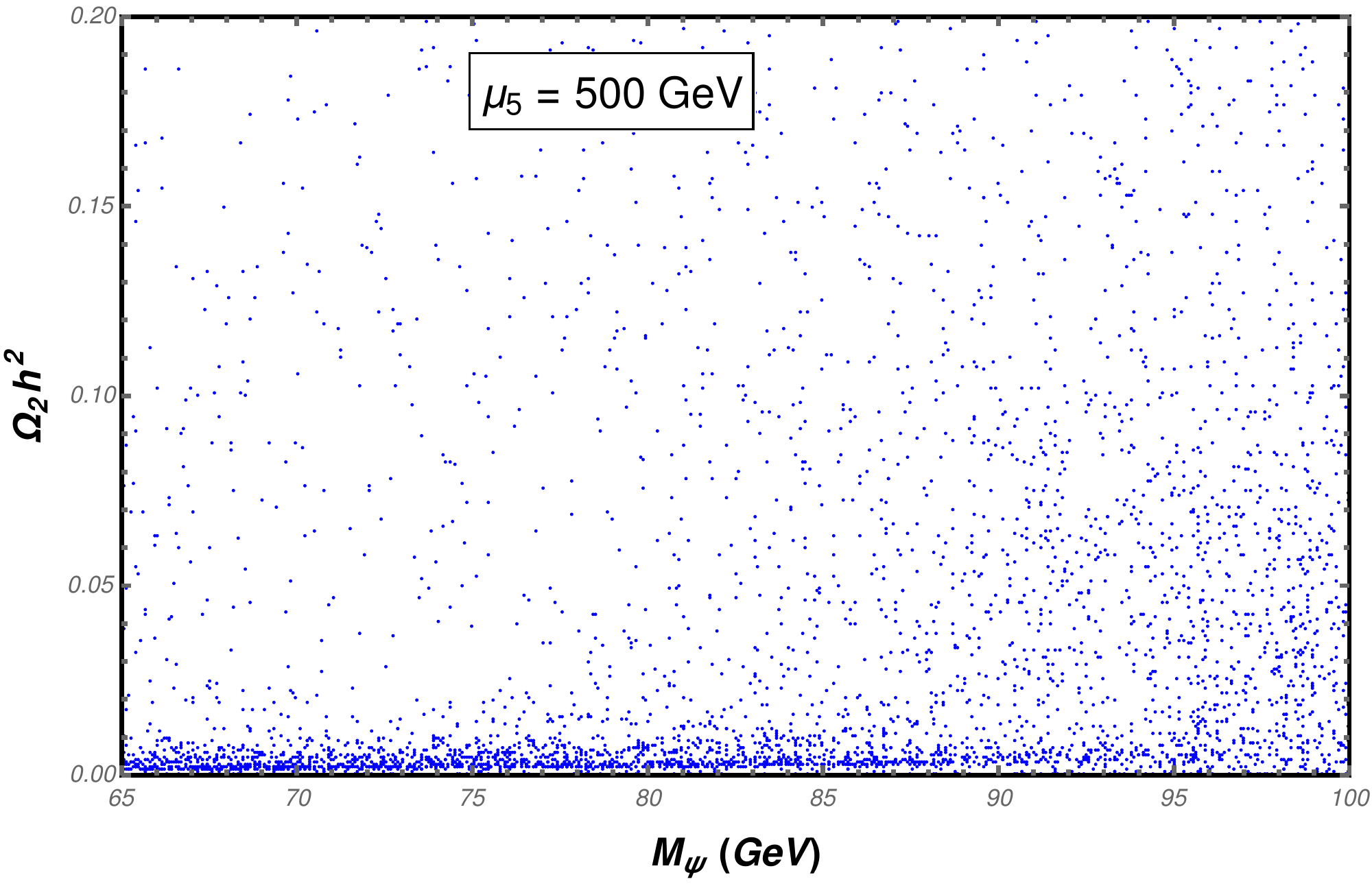}\\
  \bigskip
  \includegraphics[width=0.3\linewidth]{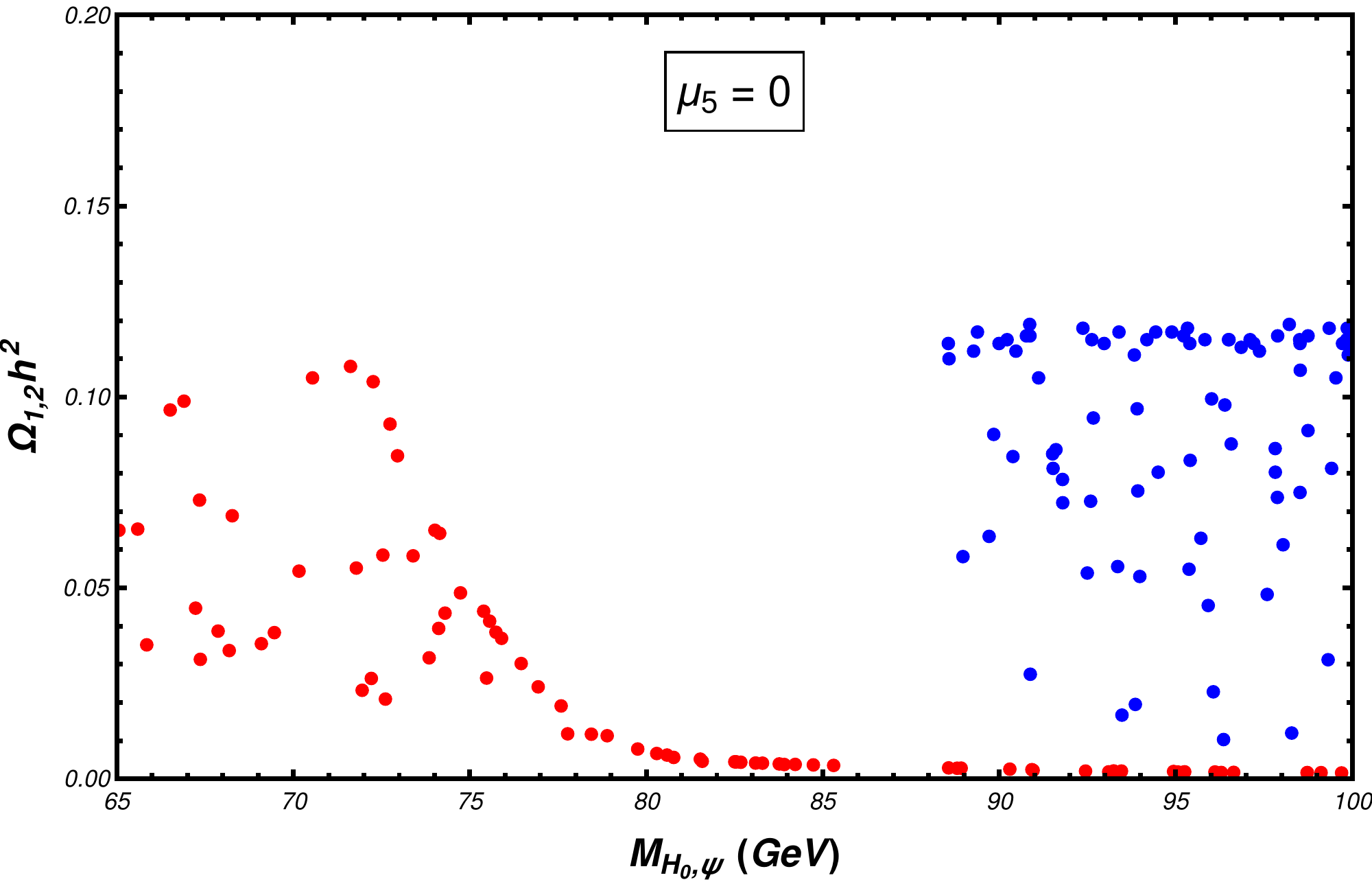}
  \hfill
  \includegraphics[width=0.3\linewidth]{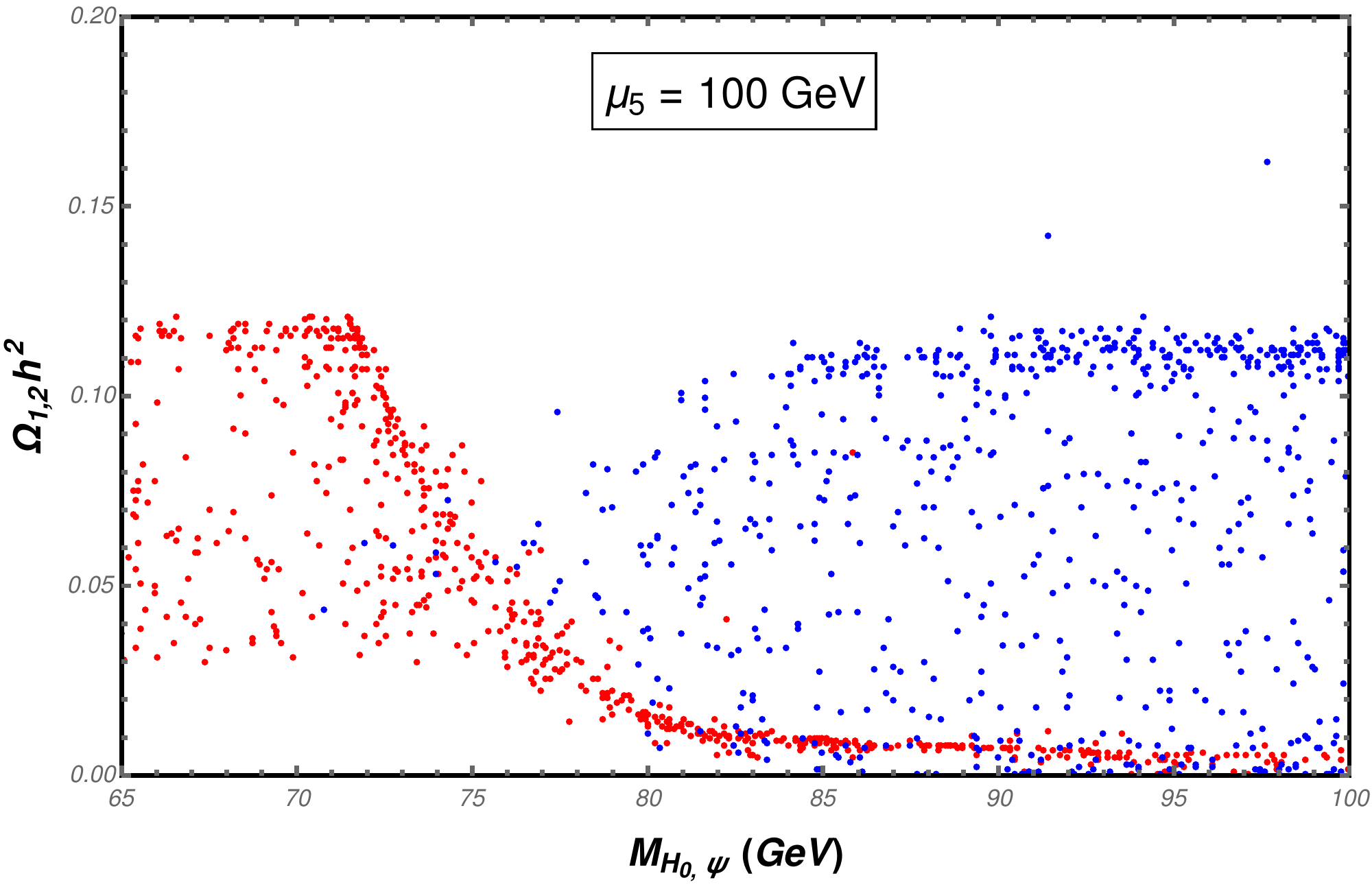}
  \hfill
  \includegraphics[width=0.3\linewidth]{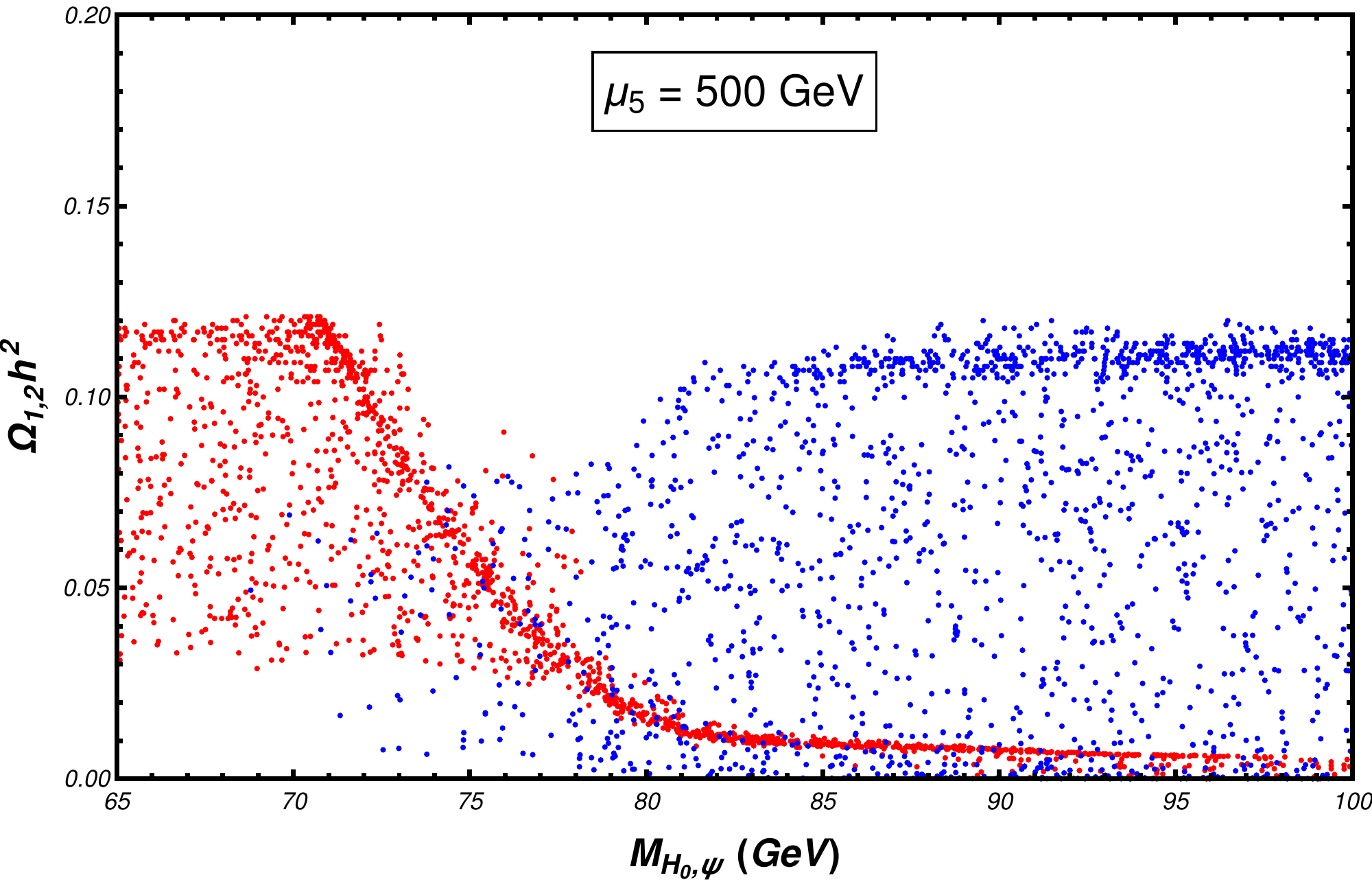}
  \caption{In {\it Top Row} $\Omega_{1}\rm h^2$ as a function of $M_{H_0}$ (GeV) and in {\it Middle Row} $\Omega_{2}\rm h^2$ as a function of $M_{\psi}$ (GeV) for illustrative values of $\mu_5$(GeV), indicating  the importance of the conversion channel ($\psi\psi \rightarrow H_0H_0$).  The total number of points generated are the same in all the cases. {\it Third Row} shows selected points from the corresponding plots in the top and middle rows compatible with $\Omega_{\rm tot}\rm h^2$ within the observed bound. {\it Red} points correspond to $\Omega_{1}\rm h^2$ vs. $M_{H_0}$, while {\it blue} points correspond to $\Omega_{2}\rm h^2$ vs. $M_{\psi}$.}
  \label{figure-omega-mH0}
  \end{figure}
  
The first row of Fig.~\ref{figure-omega-mH0} shows the variation of $\Omega_1\rm h^2$ vs $M_{H_0}$ for specific choice of $\mu_5$. In the case of $\Omega_1  \rm h^2$ with  $\mu_5=0$ dark matter annihilating into SM particles alone is possible, like in the pure IDM case. Consequently, for  $M_{H_0}\ge M_W$ the annihilation cross section controlled by gauge coupling leaves $H_0$ underabundant. However, with non-zero value of $\mu_5$ in the region $M_\psi \ge M_{H_0}$, $\psi\psi \rightarrow H_0H_0$ boosts up the relic density. This is illustrated in the top-middle plot of Fig.~\ref{figure-omega-mH0}. For very large $\mu_5$, increase in annihilation  cross section reduces the relic density below the observed limit, making the model less compatible as indicated in the top-right plot of Fig.~\ref{figure-omega-mH0}. Second row of Fig.~\ref{figure-omega-mH0} shows the variation of $\Omega_2  \rm h^2$ vs. $M_{\psi}$. In this case, $\psi$ is overabundant until the gauge boson annihilation channels open up around $M_\psi\sim M_W$, when $\mu_5$ is set to zero, as shown in the left plot. On the other hand, with $\mu_5\ne 0$, the additional channels including the annihilation into $H_0H_0,~ A_0A_0$ and $H^+H^-$ bring down the relic density to acceptable levels for $M_\psi \le M_W$ as well.

\begin{figure}[H]
\centering
  \includegraphics[width=0.45\linewidth]{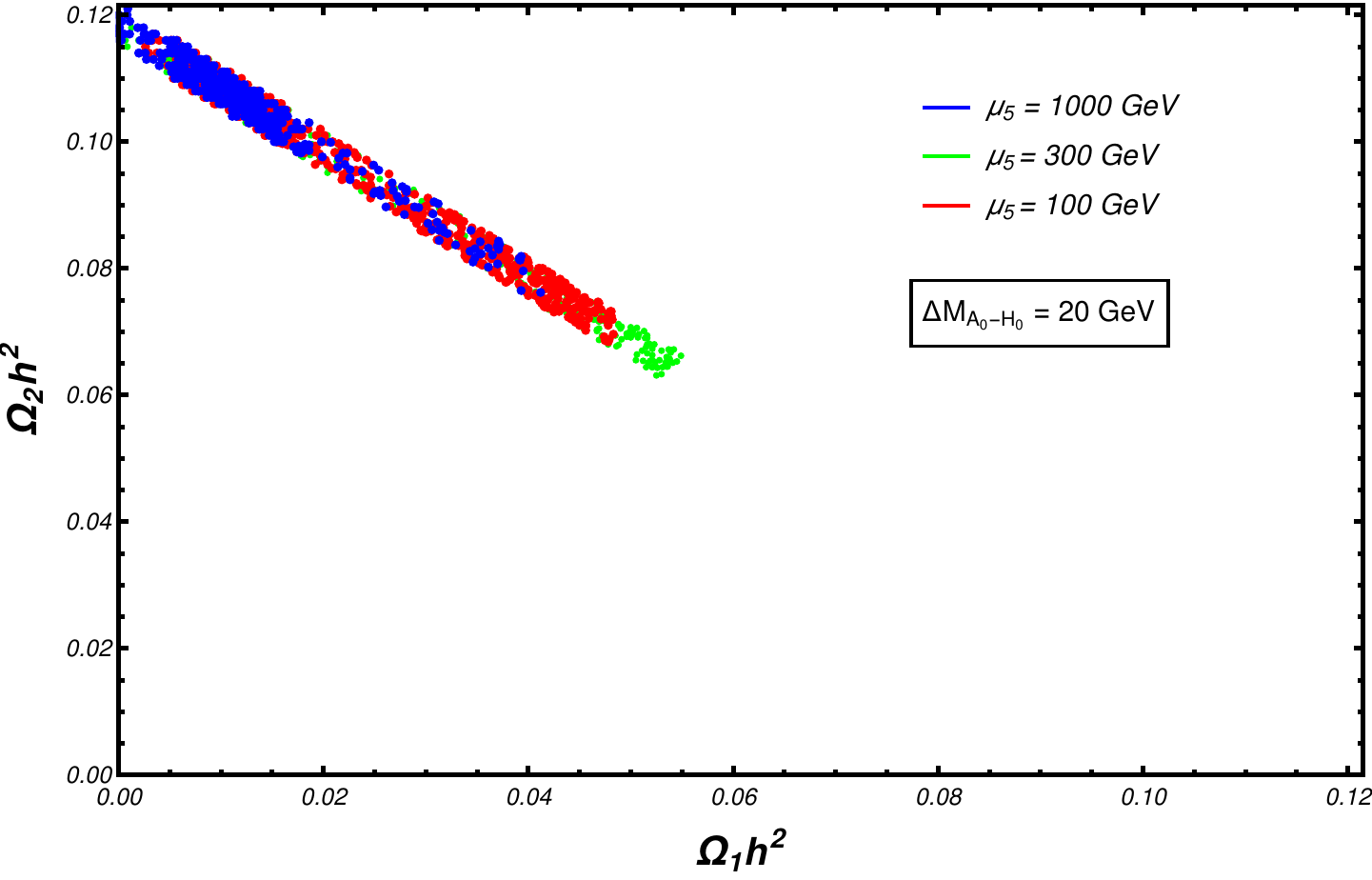}
  \hfill
  \includegraphics[width=0.45\linewidth]{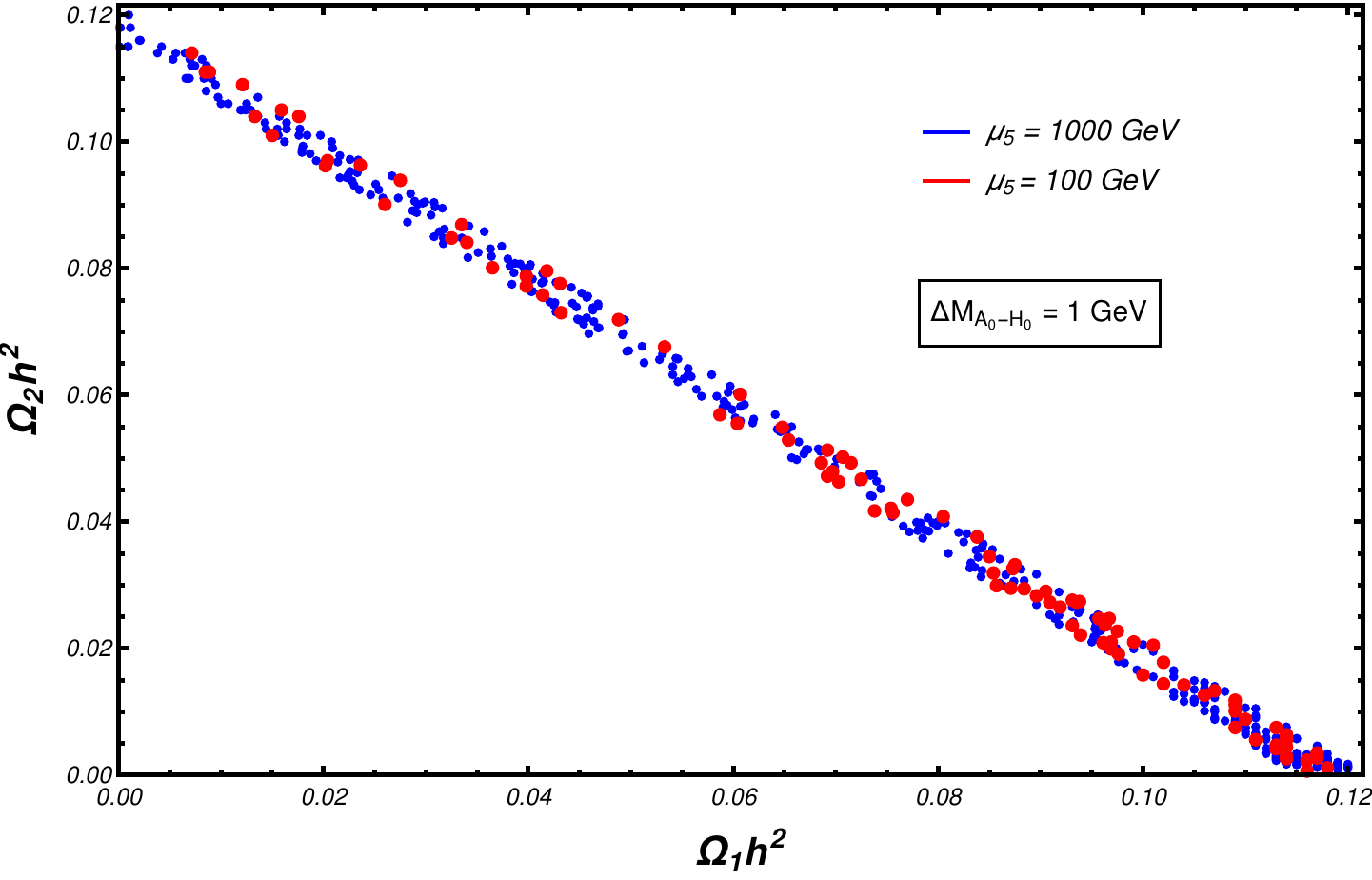}
  \caption{ Correlation between $\Omega_1  \rm h^2$ and $\Omega_2  \rm h^2$ for points satisfying total relic density bound for illustrative values of $\mu_5$. Two different mass splitting of $\Delta M_{A_0-H_0}=1$ GeV ({\it right}) and 20 GeV ({\it left}) of the IDM neutral Higgs bosons show the importance of co-annihilation channels.}
  \label{mH0mpsi-Om12}
\end{figure}

 Fig.~\ref{mH0mpsi-Om12} shows the correlation between $\Omega_1  \rm h^2$ and $\Omega_2  \rm h^2$ at a mass splitting $\Delta M_{H_0-A_0}$=1 GeV (right) and 20 GeV (left) for a mass range of $500\le M_{DM}\le 1000$ GeV.   When the mass splitting is small (of the order of 1 GeV), 
  co-annihilation between the inert scalars counters the gauge suppression and increases the scalar relic density substantially. In this case,  the effect of $\mu_5$ is negligible as seen from  Fig.~\ref{mH0mpsi-Om12} (right). 
 On the other hand, for a larger mass splitting, when the co-annihilation is suppressed, one may achieve 
 significant contribution from $\Omega_1{\rm h}^2$ for a suitably chosen value of $\mu_5$. We find that the best case scenario corresponds to a value of $\mu_5\sim 300$ GeV.
Hence in the multi-component scenario, the contribution from $\Omega_1  \rm h^2$ is boosted up compared to the single component case, thanks to the conversion from the fermionic component.  
We would like to reiterate the advantage of the multi-component case considered here, which deviate from the purely IDM like scenario, where this mass range of DM is available only for closely degenerate case of $M_{H_0}\sim M_{A_0}$.

\begin{figure}[h]
\centering
  \includegraphics[width=0.6\linewidth]{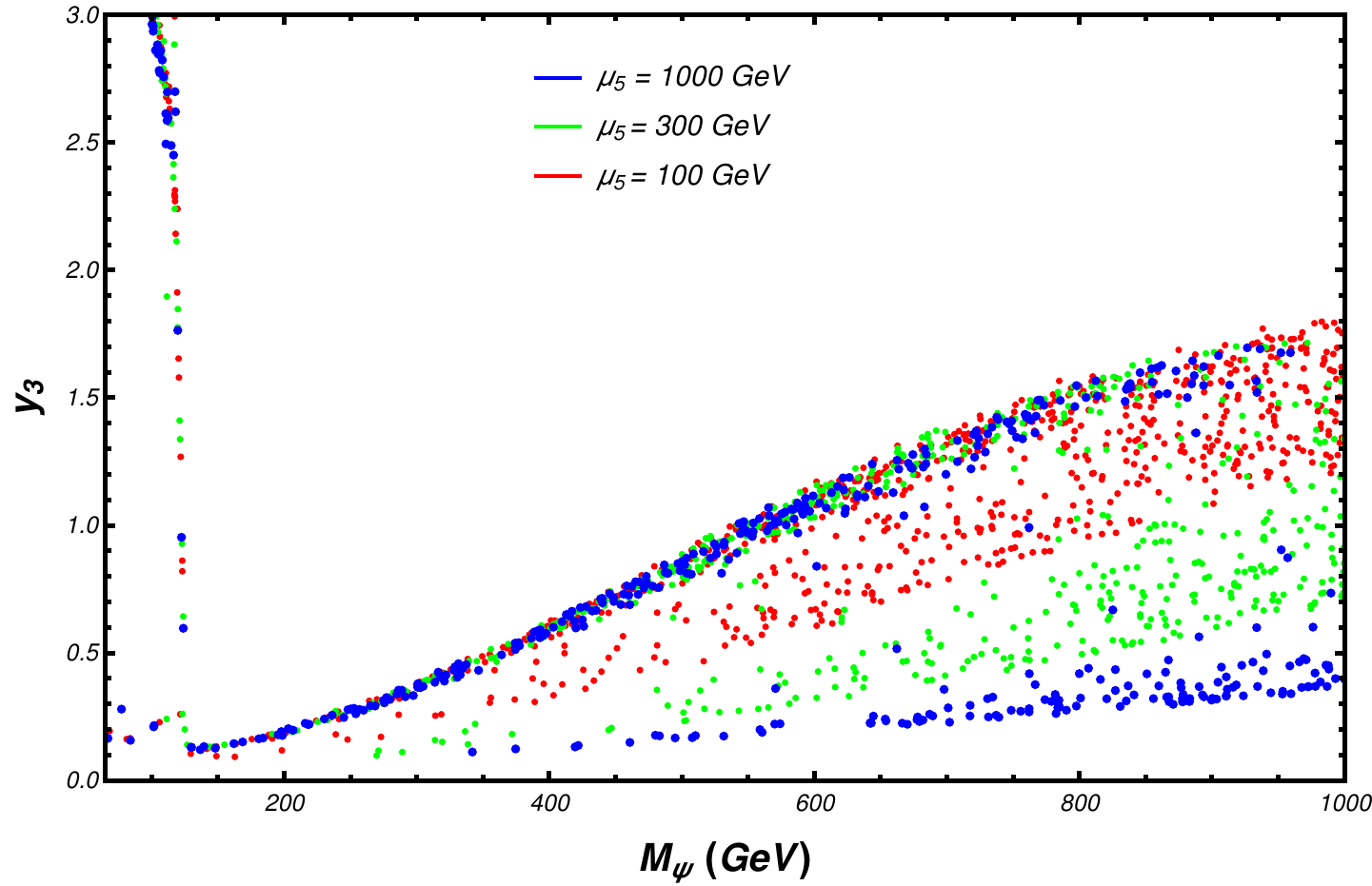}
  \caption{Regions of $y_3-M_\psi$ plane compatible with the relic density bound for specific choices of $\mu_5$ values,  indicating the strong presence of the conversion channel $\psi\psi\rightarrow H_0H_0$ in the resonant region with $M_{H_S}=2M_\psi$.}
  \label{figure_y3Mpsi_1}
\end{figure}

Presenting th correlation between $M_\psi$ and the Yukawa coupling $y_3$ between $\psi\psi H_S$ 
 Fig.~\ref{figure_y3Mpsi_1} shows that higher value of $\mu_5$ corresponds to lower $y_3$ for a fixed value of $M_{\psi}$, implying that $\psi\psi \rightarrow H_0 H_0$ is a dominant channel as per relic density consideration. Remember that $\mu_5$ corresponds to the $H_SH_0H_0$ interaction. To maintain the cross-section at a certain value in order to follow the relic bound, higher $\mu_5$ will correspond to lower $y_3$ and vice versa. All points here satisfy the total relic density bound. Notice that we have considered the resonance condition $M_{H_S}=2M_{\psi}$, while $M_{\psi}$ varies in the full range of Scenario 2,  bringing in the relevant $s$-channel annihilation of $\psi$ mediated by $H_S$.
\begin{figure}[H]
\centering
  \includegraphics[width=0.6\linewidth]{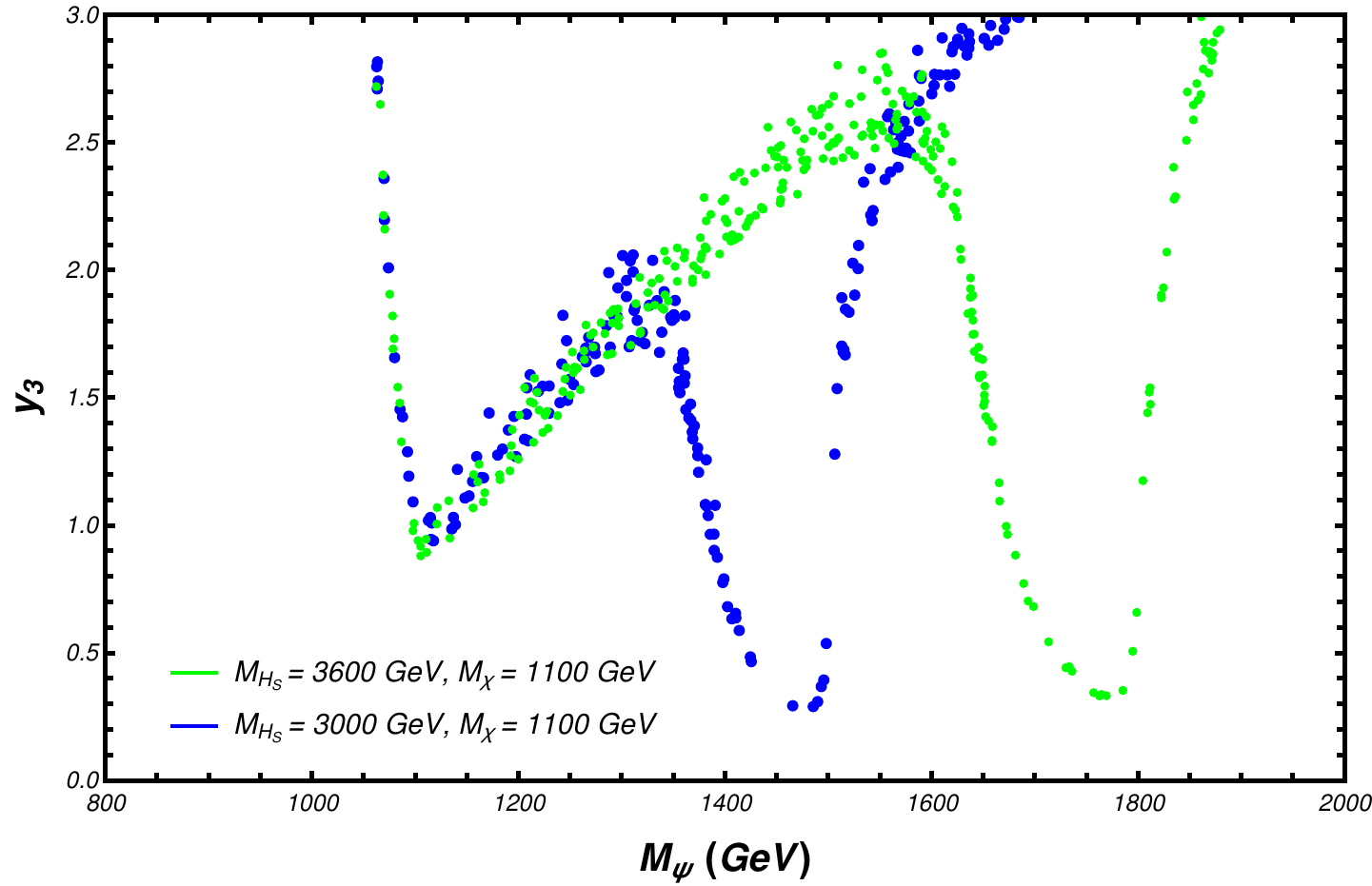}
  \caption{$y_3$ vs $M_\psi$ for fixed $M_{H_S}$ and $M_\chi$ values, showing the presence of the annihilation channel $\psi\psi\rightarrow \chi\chi$ and the resonant behaviour with $M_{H_S}=2M_\psi$.}
    \label{figure_y3Mpsi_2}
\end{figure}

In the large $M_{\psi}$ region, the DM annihilation into $\chi$ pair opens up adding further possibilities. The  $\psi\psi \rightarrow \chi\chi$ is an $s$-channel process mediated via $H_S$. Thus, the couplings $y_2$ and $y_3$ along with the masses of $H_S$, $\chi$ and $\psi$ decide the cross section.  Fig.~\ref{figure_y3Mpsi_2} shows the allowed regions of $y_3-M_\psi$ plane for specifically chosen values of the Yukawa coupling $y_2$ and the masses $M_\chi$ and $M_{H_S}$, which is set to $y_2=0.5$. The threshold is seen as $M_\psi\sim M_\chi$, which is set to 1100 GeV here. Further, the depression around the resonant condition of $M_{H_S}=2M_\psi$ is clearly seen, as expected in the $s$-channel process. Since in this region, the mass splitting between $H_0$ and $A_0$ is kept at 10 GeV, $\Omega_1\rm h^2$ is always small, hence the dominant contribution in $\Omega_{\rm tot}\rm h^2$ comes from $\Omega_2\rm h^2$. 

\section{Summary and conclusions}\label{summary}

The scalar-fermionic multipartite scenarios discussed in the literature have the DM candidates as gauge singlets  and consequently highly constrained by the direct detection experiments and relic density limits. In these scenarios, the direct detection prefers small portal couplings, which consequently provide much smaller cross section than required to contain the overabudance of the dark matter relic density. On the other hand, the gauge non-singlet scalar dark matter models like the IDM overkills the dark matter leading to underabundance except when they are close to a TeV of mass.  We have demonstrated that the presence of a fermionic partner (denoted as $\chi$) to the inert scalar would alleviate these difficulties, opening up the low mass regions as well.  
Along this line, we present a novel scenario with possibility of scalar and fermionic dark matter coexisting, compatible with all the experimental bounds including the relic density measurements, the direct detection limits and the collider constraints. We consider the presence of a gauge singlet fermion interacting with visible sector through Higgs portal couplings, along with the gauge doublet scalar dark matter. A wide range of parameter space(10 GeV - 2 TeV) for dark matter mass is considered and the possible  signatures are analyzed. We find that, the entire mass range is compatible with the relic density and direct detection bounds. 
The dark matter particles interact among themselves opening possibilities of conversion from one type to the other leading to interesting phenomenology and compensates for the underabundance of the individual relic density of $H_0$ in the otherwise not compatible range of 80 GeV $\lesssim M_{H_0} \lesssim 500$ GeV. For the entire mass range of the inert scalar $H_0$, the fermionic dark matter candidate $\psi$ contributes to the observed relic density starting from a few GeV to the TeV range.
The lepton portal annihilation channels contribute to the relic density of $H_0$, denoted by $\Omega_1\rm h^2$, without adding to the direct detection cross-section, being a $t$-channel processes mediated by the fermionic partner $\chi$. Hence keeping $\lambda_L$ fixed at an admissible low value at par with the direct detection limits, the lepton portal couplings and mass of $\chi$ can be adjusted to get the correct relic density for $H_0$. 
Owing to the conversion of fermionic dark matter pair into a scalar dark matter pair substantial contribution of 
 $\Omega_1\rm h^2$ in the total relic density is possible. The fermionic component can suffice for the deficit in the total relic density as well. On the other hand,  for large rate of annihilation of $H_0$ into SM  through the Higgs portal channels, $\Omega_1\rm h^2$ could become very small.  In such scenarios, the fermionic component dominates the scene with the dark matter scenario effectively becoming a single-component case. In the large mass region (500 GeV - 2 TeV) typical IDM contributes substantially to the relic density for very low mass splitting between the inert scalars, thanks to the now relevant co-annihilation channels. However, the effect of co-annihilation is negligible with larger mass splitting making it non-compatible with the relic density measurements. In the model discussed here, fermionic to scalar dark matter conversion permits even a larger mass splitting to produce the correct relic density. 

Finally, we expect that the model can bring in interesting collider phenomenology with the fermionic partner, $\chi$ of the inert scalar doublet changing the production and decay patterns of the IDM charged scalars in the mass range that could be probed at LHC. We defer a detailed collider study for a future work.

\vskip 5mm
\noindent{\large \bf Acknowledgement}\\[2mm]
The authors would like to acknowledge DST-SERB, India research grant EMR/2015/000333, and the DST-FIST grant SR/FST/PSIl-020/2009 for the computing facility for partly supporting this work. SC thanks Rashidul Islam for fruitful discussions and acknowledges the Ministry of Human Resource Development, India, for the Research Fellowship.

\appendix
\section{Expressions for Cross sections}
\setcounter{equation}{0}
\renewcommand{\theequation}{\thesection\arabic{equation}}
Cross section for $H_0 H_0\rightarrow \tau^+ \tau^-$, neglecting the $\tau$ mass, is given by
\begin{eqnarray}
\sigma_{H_0 H_0\rightarrow \tau^+ \tau^-}&=&\frac{1}{16 \pi s^2}\frac{1}{\beta_{H_0}^2}\int_{t_{min}}^{t_{max}} \overline{|{\cal M}^2_{H_0}|}~ dt
\label{h0h0tautau}
\end{eqnarray}
where $t_{min}=-\frac{s}{4}\left(1+\beta_{H_0}\right)^2$, $t_{max}=-\frac{s}{4}\left(1-\beta_{H_0}\right)^2$ with  $\beta_{H_0}=\sqrt{1-\frac{4M_{H_0}^2}{s}}$, and the square of the invariant amplitude ,
\begin{equation}
\overline{|{\cal M}^2_{H_0}|}= \frac{y_1^4}{4}\left[M_{H_0}^4+t(s-M_{H_0}^2)+t^2\right]~\left\{\frac{2}{(u-M_{\chi}^2)(t-M_{\chi}^2)}-\frac{1}{(u-M_{\chi}^2)^2}-\frac{1}{(t-M_{\chi}^2)^2}\right\} .\nonumber 
\end{equation}
The fermionic dark matter $\psi$ can annihilate into the scalar dark matter $H_0$ when kinematically favoured. Such conversion are relevant in situations where $H_0$ is underabundant, as in Scenario 2 discussed in Section \ref{sec_ov_largemass}.
\begin{eqnarray}
\sigma_{\psi \bar{\psi}\rightarrow H_0 H_0}&=&\frac{y_3^2}{128 \pi}~\frac{\beta_{\psi}\beta_{H_0}}{s}~\left\{\frac{\lambda_{H}^2~\sin^2\alpha}{(1-\tau_H)^2+\omega^2_H}+\frac{\lambda_{H_S}^2~\cos^2\alpha}{(1-\tau_{H_S})^2+\omega^2_{H_S}}~~~~~~~~~~~~~~~~~~~~~~~\right. \nonumber \\[3mm]
&&\left.+\lambda_{H_S} \lambda_{H} ~\sin2\alpha~\frac{(1-\tau_{H_S})(1-\tau_H)+\omega_H\omega_{H_S}}{\left[(1-\tau_H)^2+\omega^2_H\right]~\left[(1-\tau_{H_S})^2+\omega^2_{H_S}\right]}\right\}
\label{psipsih0h0}
\end{eqnarray}
\vskip 5mm
\noindent
where $\lambda_{H}=\frac{1}{\sqrt{2}}\left(\mu_5 \sin\alpha-2\sqrt{2}~ \lambda_L v~ \cos\alpha\right),~ \lambda_{H_S}=\frac{1}{\sqrt{2}}\left(\mu_5 \cos\alpha+2\sqrt{2}~ \lambda_L v~ \sin\alpha\right)$,  $\tau_i=\frac{M_i^2}{s}$,  $\omega_i=\frac{\Gamma_iM_i}{s}$ and $\beta_i=\sqrt{1-\frac{4m_i^2}{s}}$. \\

The cross sections for $\psi$ pair annihilation into SM particles relevant to discussion in Section \ref{sec_ov_largemass} are given below. 

\begin{eqnarray}
\sigma_{\psi \bar{\psi}\rightarrow WW} = \frac{ y_3^2}{64 \pi}\frac{m_W^4}{ v^2}~\frac{\beta_{\psi}\beta_W}{s}~ \left(3-\frac{s}{M_W^2}+\frac{s^2}{4 M_W^4}\right)~\sin^22\alpha~
\left\{\frac{(\tau_H-\tau_{H_S})^2+(\omega_H-\omega_{H_S})^2}
{\left[(1-\tau_H)^2+\omega^2_H\right]~\left[(1-\tau_{H_S})^2+\omega^2_{H_S}\right]}\right\},~~~~~
\label{psipsiWW}
\end{eqnarray}

\begin{eqnarray}
\sigma_{\psi \bar{\psi}\rightarrow ZZ} = \frac{ y_3^2}{128 \pi}\frac{m_Z^4}{ v^2}~\frac{\beta_{\psi}\beta_Z}{s}~ \left(3-\frac{s}{M_Z^2}+\frac{s^2}{4 M_Z^4}\right)~\sin^22\alpha~
\left\{\frac{(\tau_H-\tau_{H_S})^2+(\omega_H-\omega_{H_S})^2}
{\left[(1-\tau_H)^2+\omega^2_H\right]~\left[(1-\tau_{H_S})^2+\omega^2_{H_S}\right]}\right\},~~~~~
\label{psipsiZZ}
\end{eqnarray}

\begin{equation}
\sigma_{\psi \bar{\psi}\rightarrow f \bar{f}} =N_C \frac{ y_3^2}{8 \pi}\frac{m_f^2}{ v^2}~\frac{\beta_{\psi}\beta_f^3}{s}~ \sin^22\alpha~
\left\{\frac{(\tau_H-\tau_{H_S})^2+(\omega_H-\omega_{H_S})^2}
{\left[(1-\tau_H)^2+\omega^2_H\right]~\left[(1-\tau_{H_S})^2+\omega^2_{H_S}\right]}\right\},~~~~~
\label{psipsiff}
\end{equation}
where $N_C=1$ for leptons and $3$ for quarks.

\bigskip 
Cross section for $\psi$ pair annihilating into $H_S$ pair is given by
\small{\begin{eqnarray}
\sigma_{\psi \bar{\psi}\rightarrow H_S H_S}&=&\frac{1}{16 \pi s^2}\frac{1}{\beta_{\psi}^2}\int_{t_{min}}^{t_{max}}
\overline{|{\cal M}^2_{H_S}|} ~dt 
\label{psipsihshs}
\end{eqnarray}}
where $t_{min}=-\frac{s}{4}(\beta_{H_S}+\beta_\psi)^2$, $t_{max}=-\frac{s}{4}(\beta_{H_S}-\beta_\psi)^2$ and the invariant amplitude square is
\begin{eqnarray}
\overline{|{\cal M}^2_{H_S}|}&=& \frac{y_3^4 \cos^4\alpha}{8}\left[\frac{M_{H_S}^4+2 M_{H_S}^2(M_{\psi}^2-t)-(4 M_{\psi}^2-s-t)(3 M_{\psi}^2+t)}{(t-M_{\psi}^2)(u-M_{\psi}^2)}\right. \nonumber \\
&&\left. -\frac{t^2+(s-2 M_{H_S}^2)t-M_{\psi}^2(s-7t+6 M_{H_S}^2)+M_{H_S}^4+8 M_{\psi}^4}{(t-M_{\psi}^2)^2}\right] \nonumber
\end{eqnarray}
\bigskip

Cross section for $\psi$ annihilating into a pair of $\chi$ is 
\small{\begin{eqnarray}
\sigma_{\psi \bar{\psi}\rightarrow \chi^+ \chi^-}&=&\frac{y_3^2 y_2^2}{64 \pi s}~\beta_{\psi}\beta_{\chi}^3~\left\{\frac{\sin^4\alpha}{(1-\tau_H)^2+\omega_H^2}+\frac{\cos^4\alpha}{(1-\tau_{H_S})^2+\omega_{H_S}^2}\nonumber \right.\\ 
&&\left. +2\sin^2 \alpha\cos^2\alpha~\frac{\omega_H\omega_{H_S}+(1-\tau_H)^2(1-\tau_{H_S})^2}{[(1-\tau_H)^2+\omega_H^2]~[(1-\tau_{H_S})^2+\omega_{H_S}^2]}
\right\}
\label{psipsichichi}
\end{eqnarray}}

\bibliographystyle{JHEP}
\bibliography{MCDM} 
\end{document}